\documentclass[journal,twoside]{IEEEtran}
\usepackage{amsmath,amsthm}
\usepackage{enumerate}
\usepackage[noadjust]{cite}
\usepackage{}
\usepackage{graphicx,epstopdf}
\usepackage{xfrac,amssymb}
\usepackage[table]{xcolor}
\usepackage[font=small,skip=0pt]{caption}
\usepackage[normalem]{ulem}
\usepackage{comment}
\usepackage{color}
\definecolor{dgreen}{RGB}{73,153,0}
\usepackage{multirow}
\usepackage{mathtools}
\usepackage{caption}
\usepackage{subcaption}

\graphicspath{{Figures/}} 

\usepackage{mdframed}

\usepackage{algorithm}
\usepackage[noend]{algpseudocode}
\algnewcommand{\Initialize}[1]{%
  \State \textbf{Initialize:}
  \Statex \hspace*{\algorithmicindent}\parbox[t]{.8\linewidth}{\raggedright #1}
}

\newcommand{\argmax}[1]{\underset{#1}{\operatorname{arg}\,\operatorname{max}}\;}

\begin{document}
%
\title{Vector Quantization Methods for Access Point Placement in Cell-Free Massive MIMO Systems}
%
%
%

\author{Govind~R.~Gopal,~\IEEEmembership{Graduate~Student~Member,~IEEE}~and~Bhaskar~D.~Rao,~\IEEEmembership{Fellow,~IEEE}
\thanks{Govind R. Gopal and Bhaskar D. Rao are with the Department
of Electrical and Computer Engineering, University of California San Diego, La Jolla,
CA 92093, USA (e-mail: ggopal@ucsd.edu; brao@ucsd.edu).
The work of Govind R. Gopal was supported in part by the Center for Wireless Communications (CWC), University of California San Diego and in part by Qualcomm Inc. through the Faculty-Mentor-Advisor program. The work of Bhaskar D. Rao was supported by the National Science Foundation (NSF) under Grant CCF-2124929 and Grant CCF-2225617.
}
%
}


\maketitle
\begin{abstract}
We examine the problem of uplink cell-free access point (AP) placement in the context of optimal throughput. In this regard, we formulate two main placement problems, namely the sum rate and minimum rate maximization problems, and discuss the challenges associated with solving the underlying optimization problem with the help of some simple scenarios. 
As a practical solution to the AP placement problem, we suggest a
vector quantization (VQ) approach.
The suitability of the VQ approach to cell-free AP placement is investigated by examining three VQ-based solutions. First, the standard VQ approach, that is the Lloyd algorithm (using the squared error distortion function) is described. Second, the tree-structured VQ (TSVQ), which performs successive partitioning of the distribution space is applied. Third, a probability density function optimized VQ (PDFVQ) procedure is outlined, which enables efficient, low complexity, and scalable placement, and is aimed at a massive distributed multiple-input-multiple-output (MIMO) scenario. While the VQ-based solutions do not solve the cell-free AP placement problems explicitly, numerical experiments show that their sum and minimum rate performances are good enough, and offer a good starting point for gradient-based optimization methods. 
Among the VQ solutions, PDFVQ, with advantages over the other VQ methods, offers a good trade-off between sum and minimum rates.
\end{abstract}

\begin{IEEEkeywords}
Base station placement, Beyond 5G, Lloyd algorithm, scalability, throughput optimization.
\end{IEEEkeywords}

%
\IEEEpeerreviewmaketitle

\section{Introduction}
\label{sec:intro}


\IEEEPARstart{T}{he} concept of massive multiple-input-multiple-output (MIMO) \cite{mar10} has emerged in recent decades as a strong solution for 5G and Beyond wireless communication systems \cite{and14,cha20,jeo21}. By having a large number of antennas, such systems enable higher spectral and energy efficiencies, and reduced interference due to increased diversity \cite{lar14,llu14}.
Distributed MIMO, comprising distributed antenna systems (DASs), offer higher average rates over their colocated counterparts \cite{cho07,tru13,wan12,koy17g}.
Distributed massive MIMO can be predominantly split into non-cooperative and cooperative systems. Small-cell systems comprise non-cooperative systems while cell-free systems constitute the latter.

Cell-free massive MIMO, where there are a very large number of antennas or access points (APs) serving users that are not divided into cells, have become popular of late as they help in mitigating interference among users and in enhanced spectral efficiency over small-cell systems \cite{nay15,ngo17,ibr17,bjo20}. The concept of ``cell-free" itself arises from network MIMO \cite{ven07}. However, it is to be noted that these benefits come at the cost of large backhaul requirements necessary for the information exchange between APs \cite{ges10,irm11} and the network controller (NC) or the central processing unit, where all or most processing is performed. The computational and processing requirements of a cell-free system are thus high, and are expected to grow with the anticipated network densification in Beyond 5G/6G wireless systems \cite{wal20}. Hence, the conventional small-cell architecture is still very much in use in current and near future 5G deployments \cite{Ghosh2019,Khorov2019}.
Nevertheless, a multitude of problems has been explored based on the cell-free network. 
Prior works investigate topics such as power optimization and energy efficiency \cite{nay17,ngo18,alo19,raj21}, rate maximization \cite{bas19,far21}, clustering (user- and cell-centric) \cite{alo17,buz17,int19,buz20,ale21}, limited fronthaul \cite{bas21}, pilot assignment \cite{buz21}, reconfigurable intelligent surfaces \cite{zha21}, and federated learning \cite{vvu21}.


%
In this work, our focus is on AP placement, with the prevailing question: \emph{How do we place APs optimally for a given user distribution?} Here, the optimality is in terms of throughput in cell-free systems.
AP placement finds application in scenarios such as stadiums, where the distribution of crowds changes according to the ongoing game, or in offices where the density of employees changes by the hour. Further, deployment of APs using unmanned aerial vehicles (UAVs) can be advantageous in areas affected by natural (or man-made) calamities where existing infrastructure has been destroyed \cite{joz14}.
Past literature has expressed interest in AP placement, however, most of the focus has been on single-cell or small-cell systems (e.g., \cite{wan09,yan15,abr18,kar21}). For instance, \cite{wan09} maximizes cell averaged ergodic capacity of a DAS by using vector quantization (VQ)-based codebook design (the Lloyd algorithm) to place antennas. In \cite{yan15}, antennas are placed circularly to optimize the average per-user rate of uniformly distributed users. In \cite{abr18}, a 10-fold improvement in capacity is shown in a distributed system over a colocated  in a simulated indoor environment. Further, aligning AP positions with the user density generated around a 40\% increase in capacity over uniformly distributed APs. The authors of \cite{kar21} and their related works study the deployment of heterogeneous wireless sensor networks (placement of both APs and fusion centers) as a source coding problem for optimal power control and with limited communication range.
Most optimization problems have mainly considered maximizing the signal-to-noise ratio (SNR) alone, but certain works have incorporated various forms of interference (e.g., two-cell leakage interference in \cite{par12} and signal-to-interference-plus-noise ratio (SINR) in \cite{gop22}). Placement of UAVs as APs (UAV-APs) has also been studied (e.g., \cite{moz16,lyu17,zha20,guo20,gop21oj,zen19}).
\cite{moz16} calculates the positions and altitudes of UAV-APs to maximize ground user coverage as well as minimize interference to users. The work of \cite{lyu17} on the other hand places UAV-APs heuristically in a spiral fashion to guarantee user coverage. The authors of \cite{zha20} find the optimal placement and coverage radius of UAV-APs that serve users with time-varying positions in Wi-Fi (IEEE 802.11) networks. In \cite{guo20}, the 3-D placement of UAV-APs with directional antennas to minimize average user transmit powers is investigated. A hybrid UAV-terrestrial AP network is considered in \cite{gop21oj} where position-flexible UAV-APs are deployed in an area occupied by fixed terrestrial APs to maximize throughput while minimizing inter-cell interference (ICI) by using a Lloyd-type algorithm.
Finally, \cite{zen19} provides an extensive tutorial on UAVs as APs and as users (cellular connected UAVs).

The problem of AP placement in cell-free systems is fairly novel and much prior research has not been conducted.
As an example, \cite{dia21} investigates the deployment of UAVs in a cell-free network to maximize minimum SINR using a gradient approach while considering pilot contamination and that not every AP communicates with all users. The authors of \cite{zhu22} consider the placement of APs in a distributed massive MIMO system as a combinatorial problem to minimize transmit powers while considering antenna radiation patterns and different channel models. While a cell-free system is not explicitly defined, the system model considered mimics such a system. A graph-based approach is found to yield significant power savings while ensuring placement with good coverage. In \cite{wan22}, 3-D placement of UAV-APs is considered to maximize the downlink sum rate and uses an alternating optimization method.
A prior work by our group \cite{nay18a} examines both sum rate and minimum rate maximization problems. This work solves the two problems using compressed sensing techniques by dividing the geographical area into regular grids. However, the approximations used do not solve the two placement problems optimally.

None of the abovementioned works have investigated VQ approaches to solve the cell-free AP placement problem. In our prior work \cite{gop21}, we preliminarily investigated the utilization of the standard VQ approach, namely the Lloyd algorithm with the squared Euclidean distance as the distortion function, to place APs in a cell-free network. The VQ technique considers a single user that communicates to its nearest AP with the objective function that utilizes a distortion function averaged over the random position of this user. It does not match the cell-free model where all users communicate to all APs. In spite of these limitations, VQ-based solutions have some features that make them worthy of consideration. By design, VQ offers not only a distributed solution but can encourage cooperation (as is expected in the cell-free model) by placing APs closer to one another in areas of higher user density. Additionally, VQ solutions provide good initial points for gradient and learning-based methods to solve specific throughput problems. Hence, in this work, we explore and compare multiple VQ techniques, each with its own benefits, that can solve for AP locations in a cell-free network, with throughput as the performance measure.

\subsection*{Contributions}
To the best of our knowledge, analysis of the cell-free AP placement problem in the context of throughput optimality and the suitability of VQ techniques to solve the same have not been addressed in past literature.
Hence, in this work, our contributions are as follows.
\begin{itemize}
    \item We formulate the two main throughput optimal cell-free AP placement problems, namely the sum rate and minimum rate maximization problems. Starting from the simpler sum SNR problem, analysis of the sum rate problem is conducted and simple examples are shown to describe the possible solutions and to highlight the challenges associated with the general AP placement problem. The minimum rate problem is briefly discussed, also with some examples.
    \item Three VQ-based techniques to place APs are proposed and explored, namely standard VQ, which is the Lloyd algorithm, tree-structured VQ (TSVQ), and probability density function optimized VQ (PDFVQ). While the Lloyd algorithm provides a well-established method to place APs, there are the disadvantage of complexity and scalability. TSVQ, through successive partitioning of the user area, places APs in such a way so as to foster cooperation. PDFVQ on the other hand, allows an efficient, less computationally intensive, and easily adaptable AP placement solution by using bit allocation, transform coding, and scalar quantization, and is especially suitable for a scaled network with a large number of APs.
\end{itemize}

The remainder of this manuscript is organized as follows. Section \ref{sec:sys_model} outlines the cell-free system model used, followed by Section \ref{sec:cf_app} which explores the two principal throughput optimization problems for AP placement, namely the sum rate and minimum rate maximizations. VQ-based techniques are described in Section \ref{eqn:vq_approaches}. In Section \ref{sec:sim_results}, we state the simulation methodology and results. Finally, concluding remarks are provided in Section \ref{sec:concl}.
Throughout this paper, we will use bold symbols to denote vectors, $\mathbb{E}\{\cdot\}$ is the expectation operator, $||\cdot||$ represents the $\ell_2$-norm of a vector, and all logarithms are to the base 2.

\section{System Model}
\label{sec:sys_model}

The system model outlined in \cite{nay18a}, \cite{gop21}, and \cite{nay18t} is used. 
$K$ single-antenna users are distributed over a geographical area with a probability density function $f_{\mathbf{P}}(\mathbf{p})$, with $\mathbf{p} \in \mathbb{R}^2$ as the random vector denoting the user position. $M$ single-antenna APs serve these users, where $\mathbf{q} \in \mathbb{R}^2$ is the AP location. With $m = 1,2,\ldots,M$ and $k = 1,2,\ldots,K$, a narrowband fading channel is considered with 
\begin{equation}
    g_{mk} = \sqrt{\beta_{mk}}h_{mk},
\end{equation}
where $\beta_{mk}$ and $h_{mk} \sim \mathcal{CN}(0,1)$ are the large- and small-scale fading coefficients, respectively, independent of each other and over coherent intervals.
A general expression for $\beta_{mk}$ is
\begin{equation}
    \beta_{mk} = \frac{ c z_{mk} }{ \left|\left| \mathbf{p} - \mathbf{q}_m \right|\right|^{\gamma} },
    \label{eqn:beta_expr}
\end{equation}
where $c$ is a constant, $z_{mk}$ is the shadow fading coefficient, and $\gamma$ is the pathloss exponent. All APs are connected via error-free backhaul links to the NC.
In the cell-free uplink regime, all users are served by all APs at the same time. The uplink received signal at AP $m$ is
\begin{equation}
y_m = \sum\limits_{k=1}^K \sqrt{\rho_r}g_{mk}s_k + w_m,
\label{eqn:CF_receivedsignal}
\end{equation}
where for user $k$, $\rho_r$ is the transmit power, $s_k$ is the data symbol with $\mathbb{E}\{|s_k|^2\}=1$, and $w_m \sim \mathcal{CN}(0,1)$ is the additive noise.
The received signal vector at the NC from all $M$ APs using \eqref{eqn:CF_receivedsignal} can be written as
\begin{equation}
    \mathbf{y} = \sum\limits_{k=1}^K \sqrt{\rho_r} \mathbf{g}_k s_k + \mathbf{w},
\end{equation}
where $\mathbf{y} = [y_{1},y_{2},\ldots,y_{M}]^T$, $\mathbf{g}_k = [g_{1k},g_{2k},\ldots,g_{Mk}]^T$, and $\mathbf{w} = [w_{1},w_{2},\ldots,w_{M}]^T$.
When a combiner $\mathbf{v}_k$ is used to estimate data symbols of user $k$ as $\hat{s}_k = \mathbf{v}_k^H\mathbf{y}$, the per-user achievable rate is $R_k = \mathbb{E}\{ \log(1 + \phi_k^{\mathbf{v}_k}) \}$, where the expectation is over all the small-scale and shadow fading coefficients, and the SINR \cite{Nayebi2016} is
\begin{equation}
    \phi_k^{\mathbf{v}_k} = \frac{ \rho_r \mathbf{v}_k^H\mathbf{g}_k\mathbf{g}_k^H\mathbf{v}_k }{ \mathbf{v}_k^H\mathbf{v}_k + \sum\limits_{\substack{k^\prime = 1\\
    k^\prime \neq k}}^K \rho_r \mathbf{v}_k^H\mathbf{g}_k\mathbf{g}_k^H\mathbf{v}_k }.
\end{equation}
One such combiner is the zero forcing (ZF) detector, and results in the processed signal at the NC as $\mathbf{r} = \left( \mathbf{G}^H\mathbf{G} \right)^{-1} \mathbf{G}^H \mathbf{y}$,
where $\mathbf{G}=[g_{mk}]$ is a $M \times K$ matrix consisting of the channel coefficients \cite{nay18a}. The achievable per-user SNR in this case is
\begin{equation}
    \psi_k^{\text{ZF}} = \frac{ \rho_r }{  \left[ (\mathbf{G}^H\mathbf{G})^{-1} \right]_{kk} }.
    \label{eqn:snr}
\end{equation}
Using an asymptotic approximation for the SNR as outlined in \cite{yan15,nay18a}, the per-user SNR can also be written as
\begin{equation}
	\frac{1}{M}\psi_k^{\text{ZF}} \xrightarrow[M\rightarrow\infty]{\text{a.s.}} \rho_r \overline{\beta}_{k},
	\label{eqn:cf_snr_approx}
\end{equation}
where
\begin{equation}
    \overline{\beta}_{k} \triangleq \lim_{M\rightarrow\infty} \frac{1}{M} \sum_m \beta_{mk}.
\end{equation}

\section{Throughput Formulations for the Cell-Free AP Placement Problem}
\label{sec:cf_app}


There are two main formulations for cell-free AP placement in terms of throughput optimality:
\begin{itemize}
    \item \textit{Sum rate maximization}, which involves the sum of the rates of all users, as follows
    \begin{equation}
    \argmax{\mathbf{q}_1,\mathbf{q}_2,\ldots,\mathbf{q}_M} \sum\limits_{k=1}^K \log(1 + \phi_k^{\mathbf{v}_k}).
    \label{eqn:opt_sum_rate_general}
    \end{equation}
    Note that this problem is identical to the average rate maximization problem by assuming a sample mean and taking the average over the user distribution $f_{\mathbf{P}}(\mathbf{p})$.
    \item \textit{Minimum rate maximization}, where the minimum of the rates among all of the users is maximized
    \begin{equation}
    \argmax{\mathbf{q}_1,\mathbf{q}_2,\ldots,\mathbf{q}_M} \min\limits_k \ \log(1 + \phi_k^{\mathbf{v}_k}).
    \label{eqn:opt_min_rate_general}
    \end{equation}
    The notion of fairness, which is important in a cell-free system since all users are served by all APs, is enforced by the minimum rate problem in \eqref{eqn:opt_min_rate_general} as opposed to the sum rate problem in \eqref{eqn:opt_sum_rate_general}. In practice, the 95\%-likely rate, which represents the best rate among the worst 5\% of the users, is used as a measure to evaluate the network minimum rate performance. Hence, the max-min rate is adjusted to the 95\%-likely rate, which is more robust. This is subsequently addressed in Section \ref{ssec:grad_approach}.
\end{itemize}

In the ensuing sections, we discuss the above formulations.
Our analysis of the sum rate maximization problem is preceded by the simpler sum SNR maximization problem (preliminarily discussed in \cite{gop21}).

\subsection{Sum SNR Maximization}
\label{ssec:sum_snr_maxztn}

The sum rate problem stated in \eqref{eqn:opt_sum_rate_general} is simplified to the sum SNR problem by using the ZF SNR $\psi_k^{\text{ZF}}$ from \eqref{eqn:snr} and by replacing the summation with an expectation (the factor of $1/K$ has been neglected), where the average is taken over the user position and the set $\mathcal{A} = \{z_{mk}, \forall m\}$ consisting of all shadow fading coefficients between each user and AP. 
The optimization problem is written as
\begin{equation}
    \argmax{\mathbf{q}_1,\mathbf{q}_2,\ldots,\mathbf{q}_M} \mathbb{E}_{\mathcal{A},\mathbf{p}} \left\{ \psi_k^{\text{ZF}} \right\}.
    \label{eqn:max_snr}
\end{equation}
We can simplify the above objective function using the approximation in \eqref{eqn:cf_snr_approx} and $\beta_{mk}$ from \eqref{eqn:beta_expr} in the following manner
\begin{equation}
\begin{aligned}
    \mathbb{E}_{\mathcal{A},\mathbf{p}} \left\{ \psi_k^{\text{ZF}} \right\} &= \mathbb{E}_{\mathcal{A},\mathbf{p}} \left\{ \sum\limits_{m=1}^M \beta_{mk} \right\},\\
    &= \mathbb{E}_{\mathbf{p}} \left\{ \sum\limits_{m=1}^M \mathbb{E}_{\mathcal{A}} \left\{ \beta_{mk} \right\} \right\},\\
    &\overset{(a)}{=} \mathbb{E}_{\mathbf{p}} \left\{ \sum\limits_{m=1}^M \frac{ c^\prime }{ \left|\left| \mathbf{q}_m - \mathbf{p} \right|\right|^{\gamma} } \right\},\\
    &\overset{(b)}{=} c^\prime \sum\limits_{m=1}^M \mathbb{E}_{\mathbf{p}} \left\{ \frac{ 1 }{ \left|\left| \mathbf{q}_m - \mathbf{p} \right|\right|^{\gamma} } \right\},
\end{aligned}
\label{eqn:sum_snr_simplification}
\end{equation}
where $c^\prime = c \mathbb{E}_{\mathcal{A}}\{z_{mk}\}$, $\mathbf{p}$ is the position of user $k$ in $(a)$, and $(b)$ uses the fact that the expectation is a linear operator. Note that  $c^\prime$ can be ignored as it does not effect the optimization problem.

The following observations can be made regarding the solution of the above objective function in \eqref{eqn:sum_snr_simplification}.
\begin{itemize}
    \item A colocated solution is evident since there is no dependence between the terms in the summation associated with each AP $m$ and hence, the optimization for each AP can be performed separately.
    The suggested colocated solution may not be a unique solution and multiple global and/or local maxima may exist for the optimization problem.
    The complete characterization of the solution, however, depends on two factors, namely the pathloss exponent $\gamma$ used as well as the shape of the user distribution $f_{\mathbf{P}}(\mathbf{p})$ over which the expectation is taken, e.g., uni-modal versus multi-modal density functions. 
    \item In \eqref{eqn:sum_snr_simplification}, although the norm $\left|\left| \mathbf{q}_m - \mathbf{p} \right|\right|$, i.e., the Euclidean distance is strictly convex \cite{boy04} in $\mathbf{q}_m$, its inverse is neither concave nor convex and is also undefined at $\mathbf{q}_m = \mathbf{p}$ (this can, however, be avoided by adding a small positive quantity to the denominator, which could also account for the height of the AP). The sum of the expectation of the inverse over the APs thus also is neither convex or concave. Additionally, when a uni-modal distribution is assumed, we can expect a colocated solution alone with a unique maximum. However, when a multi-modal distribution is considered, it is expected that multiple local maxima exist and distributed solutions may be obtained. This is explored in Section \ref{ssec:simple_examples}.
\end{itemize}

It is worth noting that the sum SNR problem has been previously addressed in part in \cite{nay18a}. In this work, while dividing the user area into regular grid points, the sum rate maximization problem is upperbounded to a sum SNR problem and using a compressed sensing framework, approximated to a linear program (called the max-sum algorithm). However, as is expected from a sum SNR problem (as discussed above), most APs are concentrated around high user density regions (a near-colocated solution). Although a high sum rate can be achieved with this solution, users far away from the APs are severely affected in terms of throughput, resulting in poor minimum rate performance. Thus, this solution is not suitable for cell-free AP placement when fairness is considered.

\subsection{Sum Rate Maximization}
\label{ssec:sum_rate_maxztn}

Returning to the sum rate maximization problem, we rewrite \eqref{eqn:opt_sum_rate_general} by using the simplifications assumed before, as follows
\begin{equation}
    \argmax{\mathbf{q}_1,\mathbf{q}_2,\ldots,\mathbf{q}_M} \mathbb{E}_{\mathcal{A},\mathbf{p}} \left\{ \log\left(1 + \psi_k^{\text{ZF}}\right) \right\},
    \label{eqn:max_rate}
\end{equation}
and the objective function, utilizing \eqref{eqn:cf_snr_approx}, can be rewritten as
\begin{equation}
    \begin{aligned}
        \mathbb{E}_{\mathcal{A},\mathbf{p}} \left\{ \log\left(1 + \psi_k^{\text{ZF}}\right) \right\} &= \mathbb{E}_{\mathcal{A},\mathbf{p}} \left\{ \log\left(1 +  \sum\limits_{m=1}^M \beta_{mk} \right) \right\}.
    \end{aligned}
    \label{eqn:sum_rate_max_beta}
\end{equation}
Similar to the SNR problem outlined before, this objective function is neither concave nor convex. However, unlike the former, the term associated with each AP $m$ in \eqref{eqn:sum_rate_max_beta} cannot be decoupled from the terms associated with the rest of the APs. Hence, for both uni-modal and multi-modal distributions, we can expect only distributed solutions that maximize the sum rate.

In summary, for both the sum SNR and sum rate problems there may be multiple local optima suggesting that the optimization problem is complex and challenging.

\subsection{Examples for Sum SNR and Sum Rate Maximizations}
\label{ssec:simple_examples}

Given the abovementioned complexity in solving the sum SNR and sum rate problems, we now attempt to understand the problems and their solutions better. For this purpose, we explore simple examples where the aforementioned two problems are solved, and where the multiple local optima are studied to develop intuition and insight.

\subsubsection{User distribution considered}
For tractability, we consider a simple 1-D scenario where users are distributed along a line, and the placement of four APs along the line to maximize both sum SNR and sum rate.
For this purpose, we assume a bi-modal distribution since it is the simplest among multi-modal distributions that can exhibit the multiple maxima as discussed in Section \ref{ssec:sum_snr_maxztn}.
With the user position denoted by $p$, the PDF of the bi-modal Gaussian considered here is
\begin{equation}
    f_P(p) = \pi_1\mathcal{N}(p|\mu_1,\sigma^2_1) + \pi_2\mathcal{N}(p|\mu_2,\sigma^2_2),
\end{equation}
where for each Gaussian $i$, $i = 1,2$, $\pi_i$ is the probability such that $\pi_1 + \pi_2 = 1$, $\mu_i$ is the mean, and $\sigma_i$ is the standard deviation. To generate different solution structures for the two optimization problems, we consider two distinct configurations of the user distribution:
\begin{enumerate}[Conf. 1:]
    \item $\pi_1 = \pi_2 = 0.5$, $\mu_1 = -3$, $\mu_2 = 3$, and $\sigma_1 = \sigma_2 = 1$.
    \item $\pi_1 = 0.35$, $\pi_2 = 0.65$, $\mu_1 = -3$, $\mu_2 = 4$, and $\sigma_1 = \sigma_2 = 1$.
\end{enumerate}
Configuration 1 is symmetric about the origin while configuration 2 is asymmetric.

\subsubsection{Definitions of SNR and rate}
The four APs have locations $q_1$, $q_2$, $q_3$, and $q_4$, and the ZF SNR for user $k$ can be calculated as
\begin{equation}
\begin{aligned}
    \psi_k^{\text{ZF}} &= \beta_{1k} + \beta_{2k} + \beta_{3k} + \beta_{4k},\\
    &= \frac{1}{(p_k - q_1)^2} + \frac{1}{(p_k - q_2)^2} + \frac{1}{(p_k - q_3)^2} + \frac{1}{(p_k - q_4)^2},
\end{aligned}
\end{equation}
where, for simplicity, the transmit power is set to unity and the definition of the large-scale fading coefficient $\beta_{mk}$ (from \eqref{eqn:beta_expr}) assumes that shadow fading is absent, the pathloss exponent is two, and the constant is set to one. These assumptions do not change the conclusions that are obtained in this section. The sum SNR and sum rate quantities are then defined as $\sum_{k=1}^K \psi_k^{\text{ZF}}$ and $\sum_{k=1}^K \log(1 + \psi_k^{\text{ZF}})$, respectively. Note that for implementation purposes, a small quantity $\epsilon$ is added to the denominator of $\beta_{ik}$, $i = 1,2,3,4$, to prevent $\psi_k^{\text{ZF}}$ from approaching infinity.
We then plot the sum SNR and sum rate for different AP locations. 

\subsubsection{AP location solutions considered}
To evaluate and understand the sum SNR and sum rate performances of the system, we study various AP placement scenarios. 
First, a colocated solution is considered where all four APs are situated at the same location. This location is found by sweeping the AP position denoted by $q$ across the span of the user locations. Second, multiple semi-distributed solutions are selected. Instead of all APs at one location, they can be allocated to each of the Gaussians and placed at their respective means. In this scenario, we consider three situations, namely when 2 APs each are placed at $\mu_1$ and $\mu_2$, $3$ APs are at $\mu_1$ and $1$ AP is at $\mu_2$, and $1$ AP is at $\mu_1$ and $3$ APs are at $\mu_2$. These three situations are termed `Distributed (2+2)', `Distributed (3+1)', and `Distributed (1+3)', respectively.
Third, a fully distributed scenario involves starting from a distributed (2+2) solution and moving the two APs within each Gaussian away from each other until the maximum is achieved. Note that this fully distributed solution represents only a local maximum. Finally, we have the solution obtained by applying the standard Lloyd algorithm to the user distribution.

\begin{figure}[t!]
	\centering
	\includegraphics [scale=0.555] {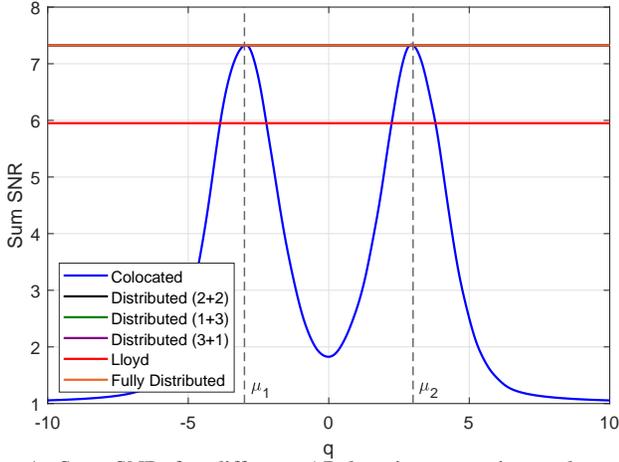}
	\caption{Sum SNR for different AP location scenarios under user configuration 1.}
	\label{fig:cf_journal_simp_eg_sum_snr_conf_1}
\end{figure}
\begin{figure}[t!]
	\centering
	\includegraphics [scale=0.555] {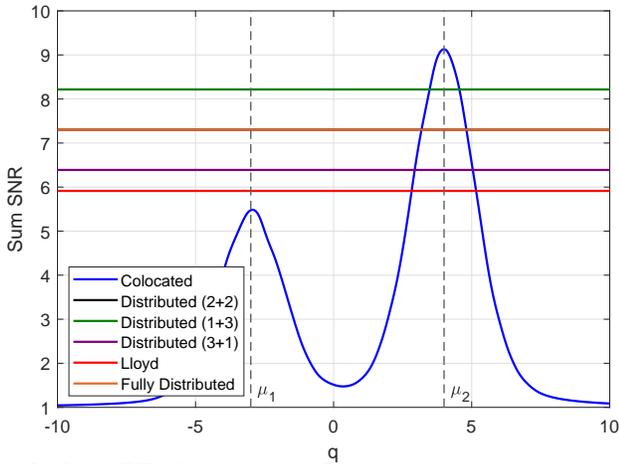}
	\caption{Sum SNR for different AP location scenarios under user configuration 2.}
	\label{fig:cf_journal_simp_eg_sum_snr_conf_2}
\end{figure}

\subsubsection{Results for sum SNR}
The results of sum SNR under user configurations 1 and 2 are shown in Fig. \ref{fig:cf_journal_simp_eg_sum_snr_conf_1} and Fig. \ref{fig:cf_journal_simp_eg_sum_snr_conf_2}, respectively. Note that for the colocated solution, the sum SNR obtained as a function of location $q$ is plotted. For the other solutions, a line is drawn corresponding to the sum SNRs obtained. 

For \textit{configuration 1}, it is observed that the peak for the colocated system occurs at the two means $\mu_1$ and $\mu_2$. The three distributed scenarios as well as the fully distributed scenario offer the same peak sum SNR value as in the colocated case. It is noted that the AP locations in the fully distributed case are the same as in distributed (2+2). Further, the Lloyd solution (which is also a fully distributed scenario) offers a lower sum SNR value. The sum SNR maximization for configuration 1 thus has multiple local maxima, including both colocated and semi-distributed solutions. 

For \textit{configuration 2}, the Gaussian with mean $\mu_2$ has a higher probability, with the result that a colocated solution with all four APs at mean $\mu_2$ yields the highest sum SNR. Among the distributed solutions, it is observed that a higher allocation of APs at $\mu_2$ favors a higher sum SNR, however, with the colocated solution offering the highest. Again, both the fully distributed and distributed (2+2) solutions have the same AP locations and sum SNR values, with the Lloyd solution performing the worst. 
In summary, the above results for sum SNR show that the AP locations that maximize the same may be colocated or distributed depending on the user distribution, as discussed in Section \ref{ssec:sum_snr_maxztn}.

\begin{figure}[t!]
	\centering
	\includegraphics [scale=0.555] {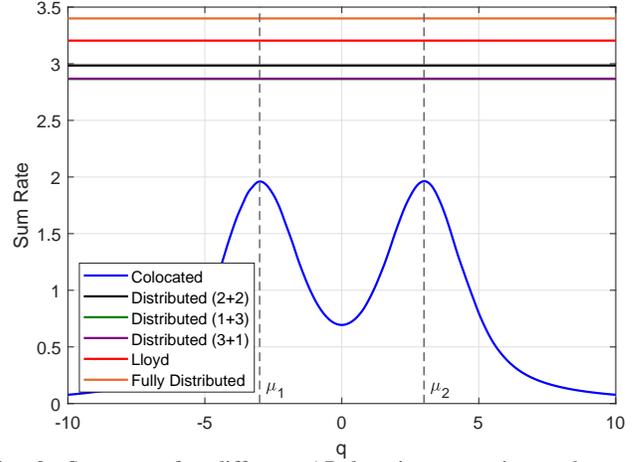}
	\caption{Sum rate for different AP location scenarios under user configuration 1.}
	\label{fig:cf_journal_simp_eg_sum_rate_conf_1}
\end{figure}
\begin{figure}[t!]
	\centering
	\includegraphics [scale=0.555] {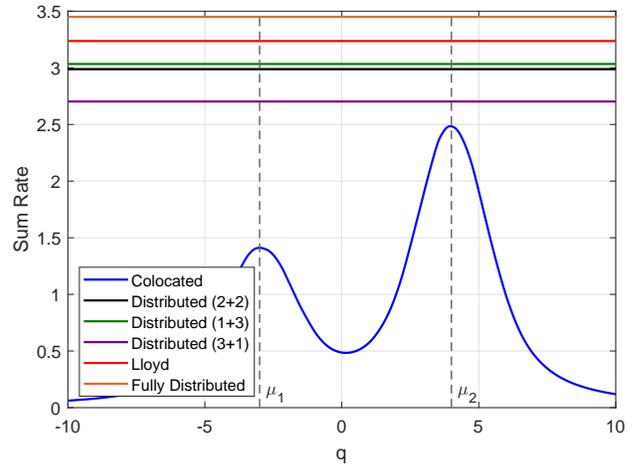}
	\caption{Sum rate for different AP location scenarios under user configuration 2.}
	\label{fig:cf_journal_simp_eg_sum_rate_conf_2}
\end{figure}

\subsubsection{Results for sum rate}
The results of the sum rate metric are shown for user configurations 1 and 2 are shown in Fig. \ref{fig:cf_journal_simp_eg_sum_rate_conf_1} and Fig. \ref{fig:cf_journal_simp_eg_sum_rate_conf_2}, respectively.
In configuration 1, it is clear that a colocated solution performs poorly and that distributed and fully distributed solutions offer a higher sum rate. Since the distribution is symmetric, both distributed (1+3) and distributed (3+1) solutions have the same sum rate, which is lower than that of distributed (2+2). The Lloyd solution performs better than the distributed solutions as no APs are colocated, and the fully distributed solution performs the best. Note that the fully distributed solution does not represent the optimum solution due to the method with which the locations are generated.
In configuration 2, the colocated solution still performs the worst. However, since the Gaussian at mean $\mu_2$ carries a higher proportion of users, the distributed (1+3) solution performs better than both distributed (2+2) and distributed (3+1). Finally, as in configuration 1, both the Lloyd and the fully distributed solution perform better than the distributed solutions.

\begin{figure}[t!]
	\centering
	\includegraphics [scale=0.555] {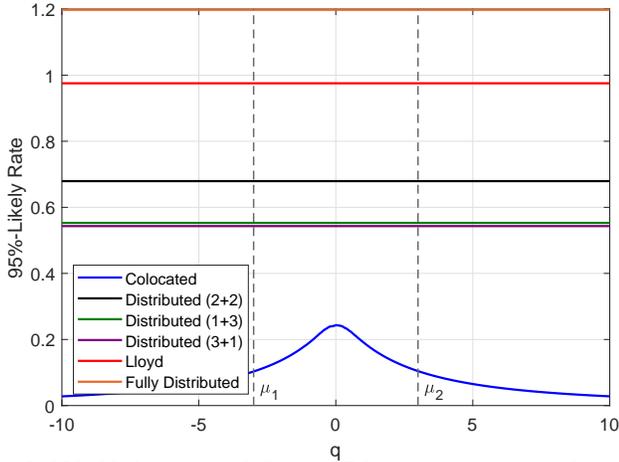}
	\caption{95\%-likely rate for different AP location scenarios under user configuration 1.}
	\label{fig:cf_journal_simp_eg_95_rate_conf_1}
\end{figure}
\begin{figure}[t!]
	\centering
	\includegraphics [scale=0.555] {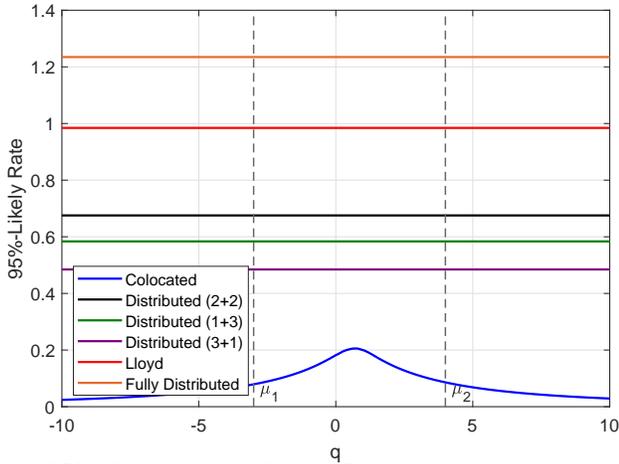}
	\caption{95\%-likely rate for different AP location scenarios under user configuration 2.}
	\label{fig:cf_journal_simp_eg_95_rate_conf_2}
\end{figure}

\subsubsection{Results for minimum rate}
We now plot the 95\%-likely rates corresponding to the different AP locations for configurations 1 and 2, in Fig. \ref{fig:cf_journal_simp_eg_95_rate_conf_1} and Fig. \ref{fig:cf_journal_simp_eg_95_rate_conf_2}, respectively.
Note that the fully distributed solution in these plots maximizes the 95\%-likely rates as opposed to the sum SNR or sum rate. 
Even though all distributed solutions generate the same peak sum SNR for configuration 1 as shown in Fig. \ref{fig:cf_journal_simp_eg_sum_snr_conf_1}, the users will not achieve the same minimum rate performance in all cases. For this user configuration, distributed (1+3) and distributed (3+1) have lower 95\%-likely rates than distributed (2+2). Additionally, the Lloyd solution has a higher minimum rate and the fully distributed solution generates the highest 95\%-likely rate. Clearly, the colocated solution exhibits the worst performance. The same observations are noted for configuration 2, with the difference being that distributed (1+3) generates a higher 95\%-likely rate than distributed (3+1) due to the higher user proportion around mean $\mu_2$.

\begin{figure}[t!]
	\centering
	\includegraphics [scale=0.555] {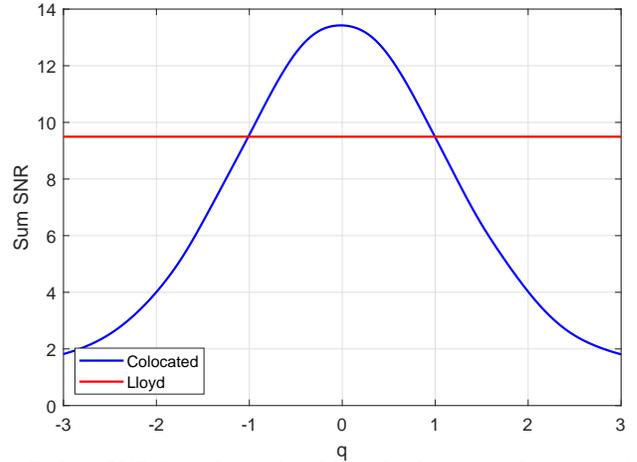}
	\caption{Sum SNR for colocated and Lloyd solutions under uni-modal distribution.}
	\label{fig:cf_journal_simp_eg_sum_snr_unimodal}
\end{figure}
\begin{figure}[t!]
	\centering
	\includegraphics [scale=0.555] {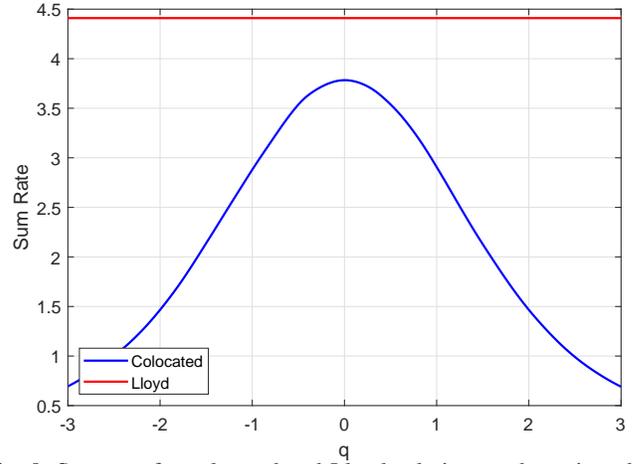}
	\caption{Sum rate for colocated and Lloyd solutions under uni-modal distribution.}
	\label{fig:cf_journal_simp_eg_sum_rate_unimodal}
\end{figure}

\subsubsection{Results for a uni-modal user distribution}
As a final note and to complete the discussion on the influence of the distribution shape on the solutions, we also plot the sum SNR and sum rate performances for a uni-modal Gaussian distribution (with zero mean and unit variance), in Fig. \ref{fig:cf_journal_simp_eg_sum_snr_unimodal} and Fig. \ref{fig:cf_journal_simp_eg_sum_rate_unimodal}, respectively.
Like the bi-modal examples shown prior, the colocated solution is preferred in the sum SNR case over the distributed Lloyd solution. The opposite is true for the sum rate case. It is also to be noted that unlike the bi-modal scenario, there is only one colocated solution that provides the peak sum SNR.

In conclusion, all observations made in the context of the simple examples support the complex nature of the AP placement for sum SNR and sum rate maximization problems (Section \ref{ssec:sum_snr_maxztn} and Section \ref{ssec:sum_rate_maxztn}).

\subsection{Minimum Rate Maximization}
\label{ssec:min_rate_max}

For the sum SNR maximization case, we showed both colocated and semi-distributed cases were favored. It was also postulated that the minimum rate performance of these solutions would be inferior to that provided by a fully distributed solution (like the Lloyd solution). Even in the sum rate maximization case, we have not addressed the minimum rates of the users explicitly and we can expect that there are users that lie far away from the AP locations, resulting in significantly low throughput to these users, which is not ideal from a fairness point of view.
The minimum rate maximization problem is thus of interest and has been defined above in \eqref{eqn:opt_min_rate_general}.
In this paper, we do not discuss this problem as it has been addressed and solved using a grid-based approach in \cite{nay18a,nay18t} and is called the max-min algorithm. Such a solution is useful since the problem can be converted into a convex problem, which can be solved easily. It is to be noted that the grid structure assumed leads to approximate solutions and a finer grid is necessary to obtain the optimal locations. In \cite{gop21}, we have further analyzed and compared the performance of the max-min algorithm with the Lloyd algorithm.

\section{Vector Quantization Approaches}
\label{eqn:vq_approaches}

In this section, as practical solutions to cell-free AP placement, we investigate how VQ techniques can be applied to sum rate and minimum rate maximizations.
%
VQ can be implemented in a cell-free system by assuming that each user is associated with its geographically nearest AP while adopting the squared Euclidean distance (error) as the distortion measure of interest.
Before we delve into the various VQ techniques, we first motivate why such an approach is useful for the placement problem.

\subsection{Why Use a VQ Approach?}

\begin{itemize}
    \item \emph{Distributedness:} The VQ approaches, through their formulations, are designed to provide a distributed solution. For example, the standard Lloyd algorithm using squared error distortion minimizes the distance of a user to its closest AP, averaged over the entire distribution. Thus, a distributed solution is obtained by ensuring that at least one AP is close to each user.
    Such a solution also addresses the minimum rate of the system in an effective manner. 
    \item \emph{Cooperation:} 
    While allowing distributedness, the VQ formulations which do not explicitly account for cooperation, also place APs close to one another. This clustering occurs especially in areas of high user density, thus encouraging cooperation that is expected in a cell-free system and addresses the system sum rate. In the case of the Lloyd algorithm, the objective function itself allows for high AP density. In TSVQ, a practical implementation of VQ discussed in Section \ref{ssec:tsvq} below, the codepoint splitting process in addition to the Lloyd algorithm used places APs closer to one another and enables more cooperation.
    \item \emph{Initialization:} Although it is possible to determine solutions for both problems through gradient methods, such methods usually generate local optima resulting in lower-than-expected performances. The VQ approaches, due to the distributed nature of their solutions, are able to provide suitable starting points for such gradient-based methods.
\end{itemize}

\subsection{Standard VQ}

The standard and simplest VQ technique is the Lloyd algorithm \cite{ger91} that utilizes the squared Euclidean distance between APs and users as the distortion measure.  This squared error (SE) distortion between a user at $\mathbf{p}$ and an AP at $\mathbf{q}_m$ is denoted as follows
\begin{equation}
d_{\text{SE}}(\mathbf{p},\mathbf{q}_m) = \left|\left| \mathbf{p}-\mathbf{q}_m \right|\right|^2.
\label{eqn:sq_error_distortion}
\end{equation}
Details of how the Lloyd algorithm can be applied to AP placement have been elucidated in \cite{gop22}. We provide the algorithm (Algorithm \ref{alg:lloyd}) below.

\begin{algorithm}
\caption{Lloyd Algorithm}\label{alg:lloyd}
\begin{algorithmic}[1]
\State Initialize random AP locations $\mathbf{q}_1^{(0)},\mathbf{q}_2^{(0)},\ldots,\mathbf{q}_M^{(0)}$.
\State Use the NNC to determine the cells $\mathcal{C}_1^{(i+1)},\mathcal{C}_2^{(i+1)},\ldots,\mathcal{C}_M^{(i+1)}$ such that
\begin{equation*}
\mathcal{C}_m^{(i+1)} \!=\! \left\{\! \mathbf{p}_k \!:\! d_{\text{SE}}\!\left(\!\mathbf{p}_k,\mathbf{q}_m^{(i)}\!\right) \!\leq \! d_{\text{SE}}\!\left(\!\mathbf{p}_k,\mathbf{q}_l^{(i)}\!\right)\!, \forall l \!\neq m \!\right\}.
\label{eqn:nnc_lloyd}
\end{equation*}
\State Use the CC to determine the AP locations $\mathbf{q}_1^{(i+1)},\mathbf{q}_2^{(i+1)},\ldots,\mathbf{q}_M^{(i+1)}$ such that
\begin{equation*}
\mathbf{q}_m^{(i+1)} = \frac{ 1 }{ \left| \mathcal{C}_m^{(i+1)} \right| } \sum\limits_{\mathbf{p}_k \in \mathcal{C}_m^{(i+1)}} \mathbf{p}_k.
\end{equation*}
\State Repeat from step 2 until convergence (MSE falls below a threshold).
\end{algorithmic}
\end{algorithm}

It is important to note that while the Lloyd algorithm provides distributed AP locations and a user is expected to be close to at least one AP, the minimum rate metric is not explicitly contained in its objective function. Additionally, the assumed cooperation between the APs enabling higher sum rate is also not explicitly included in the optimization. However, the Lloyd algorithm indirectly fosters cooperation since the AP density is proportional to a power of the user density under a high resolution approximation \cite{ger91}, as follows
\begin{equation}
    g_{\mathbf{P}}(\mathbf{p}) = \frac{ f_{\mathbf{P}}^{\frac{1}{2}}(\mathbf{p}) }{ \int f_{\mathbf{P}}^{\frac{1}{2}}(\mathbf{p}^\prime) \mathrm{d}\mathbf{p}^\prime }.
\end{equation}
Therefore, APs in the high user density areas can cooperate with one another. 

\subsection{Tree-Structured VQ}
\label{ssec:tsvq}

Tree-structured VQ (TSVQ) \cite{ger91} is an alternate VQ approach where the codebook search time is reduced compared to standard VQ.
In this technique, the input training set is partitioned into a hierarchy of Voronoi regions, which allows a tree to be generated for encoding. Thus, TSVQ differs from the standard VQ discussed above in that the final required codebook is generated by the successive splitting of intermediate codebooks, starting with a single centroid. The Lloyd algorithm (Algorithm \ref{alg:lloyd}) is applied to each stage of the hierarchy and its application is confined to the partitioned training set of the previous stage.

\emph{Why use TSVQ?} Designed primarily for relatively fast codebook search properties, TSVQ has the following benefits.
\begin{itemize}
    \item \emph{Initialization:} An advantage of TSVQ is that initialization of the AP locations is not required, since it starts with a single centroid. Compared to the standard VQ, this avoids the random initializations (considered sub-par) and the calculations needed for advanced initialization methods such as k-means++ \cite{art07}.
    \item \emph{Cooperation:} Due to the successive splitting of the intermediate codepoints, we can expect pairs of APs to be closer to each other compared to standard VQ. This enables cooperation among APs, increasing the system sum rate.
    \item \emph{Flexibility:} When the number of APs is changed for the same user distribution, codepoints can either be merged (when the number of APs is reduced) or split (when increased). The choice of codepoints for merging or splitting will be on the basis of the rate performance associated with the codepoints.
\end{itemize}
It should be noted that in general, TSVQ does not find the closest AP to each user and there is a small decrease expected in the performance of TSVQ (in terms of mean squared error) when compared to standard VQ.

\medskip\noindent
\emph{TSVQ Algorithm}. In our implementation of TSVQ for cell-free AP placement, we limit ourselves to balanced binary trees, i.e., at each stage, every intermediate codepoint is split into two codepoints, to favor the lowest complexity. The TSVQ algorithm is outlined in Algorithm \ref{alg:tsvq}. The stages are indexed by $j$ (root node is stage 0) and the set of codepoints is represented by $\mathcal{P}_j$ for stage $j$. The codepoints are indexed by $i$, and the number of codepoints increases at each stage. The codepoints generated after the algorithm converges are the required AP positions. $\mathcal{R}$ indicates the set of all users while $\mathcal{R}_i$ denotes the set of users associated with codepoint $i$. The splitting of codepoints is performed by generating two perturbations of the original codepoint.

\begin{algorithm}
\caption{Tree-Structured Vector Quantization Algorithm}\label{alg:tsvq}
\begin{algorithmic}[1]
\State Initialize $\mathcal{P}_0$ with the codepoint of all users in $\mathcal{R}$.
\State Split the codepoint(s) in $\mathcal{P}_j$ into two.
\State Apply Lloyd algorithm (Algorithm \ref{alg:lloyd}) to each split pair for training data $\mathcal{R}_i$ at stage $j$.
\State Partition $\mathcal{R}_i$ at stage $j$ into two sets corresponding to new codepoints.
\State Update $\mathcal{P}_j$ with new codepoints.
\State Repeat from step 2 for next stage $j+1$ until $M$ codepoints are generated.
\end{algorithmic}
\end{algorithm}

\subsection{PDF Optimized VQ}
\label{ssec:pdfvq}

So far, we have considered a full-scale version of VQ (the Lloyd algorithm) and TSVQ, which involves hierarchical codebook generation. Both flavors of VQ, however, come with disadvantages such as complexity, scalability, and learnability which will be elaborated shortly. In this section, we outline the probability density function optimized VQ (PDFVQ) procedure,
first developed in \cite{ana03}, for use in cell-free AP placement that can address these shortcomings. Here, an efficient quantizer is generated by first estimating the user distribution PDF using the expectation-maximization (EM) algorithm \cite{red84} and assuming a Gaussian mixture model (GMM). Then, by leveraging both bit allocation and transform coding, closed-form expressions are defined to allocate bits to each cluster and along every dimension (x- and y-coordinates of the 2-D user density) so that scalar quantizers can be used to generate the required AP codebook.
%
%

\emph{Why use PDFVQ?} Although designed for high-dimensional source inputs, this procedure addresses the shortcomings of the two prior VQ approaches in the following manner.
\begin{itemize}
    \item \emph{Complexity:} In the two prior VQ approaches, the computational complexity and memory required are high. In TSVQ, the Lloyd algorithm is applied a number of times depending on the codebook splits performed to obtain the required number of APs.
    The overall complexity of the Lloyd algorithm is $\mathcal{O}(2KMI)$, where the factor `$2$' is due to the 2-D user density and $I$ denotes the number of iterations required for convergence. It is also observed that the number of executed iterations $I$ increases with the number of APs $M$.
    In addition to the simple expressions for AP allocations for each cluster and dimension of the 2-D user density, through the use of scalar quantizers for each dimension, PDFVQ enjoys lower complexity as the quantizers work on both a lower number of users and APs, consequently performing a lower number of iterations. Further, the scalar quantizers can also be implemented by means of look-up tables \cite{jai89} that determine the codebook. Thus, there exists a simple closed-form mapping from user density to the codepoints (AP locations).
    The memory requirements in PDFVQ are also small as a result of transform coding.
    \item \emph{Scalability:} The two prior VQ solutions are not easily scalable with bit rate, that is, the number of APs. For the Lloyd algorithm alone, the whole procedure must be repeated. In the case of TSVQ, additional successive splitting steps must be performed, with the number of times that the Lloyd algorithm is implemented doubling at every step.
    In PDFVQ, when the number of bits is changed, then the PDF estimation phase remains unchanged and the bit allocation and codebook generation are readily obtained as closed-form expressions exist. Thus, PDFVQ facilitates scalability, both with respect to dimensions (of the source distribution) as well as the number of bits.
    \item \emph{Learnability:} It is expected that the user density varies over time, in which case the AP positions must be adapted to account for this change. Both prior VQ schemes are not amenable to a learning environment as they cannot adapt quickly to changes in the user distribution as well as number of APs, and their computationally complex procedures must be repeated to optimize to the new environment.
    In PDFVQ, when the source distribution changes, both the density estimation and codebook generation steps must be implemented, however, only the parameters of the density have to be learned (the EM algorithm is easily updated using existing values). If the number of APs only is changed, then the closed-form expressions (which are not altered) and scalar quantizers alone are needed, resulting in computational savings, faster adaptability, and enabling an `on-the-fly' placement.
\end{itemize}
Due to the abovementioned advantages, PDFVQ is thus suitable for AP placement both in an environment where the user distribution changes over time as well as the large-scale massive MIMO scenario with a very high number of APs. It is also worth noting that the PDFVQ procedure is independent of the chosen distortion measure for the scalar quantizers, but for simplicity, we will consider that the Lloyd algorithm is used.

\medskip\noindent
\emph{PDFVQ Algorithm}. The PDFVQ procedure from \cite{ana03} applied to AP placement (involving only two dimensions) is outlined in Algorithm \ref{alg:pdfvq} below. Note that in this paper, we model the user distributions as GMMs, which avoids the need for parameter estimation. The GMM considered is of the form
\begin{equation}
f_{\mathbf{P}}(\mathbf{p}) = \sum\limits_{l=1}^L p_l\mathcal{N}\left(\mathbf{p}\vert\boldsymbol{\mu}_l,\boldsymbol{\Sigma}_l\right),
\label{eqn:gmm}
\end{equation}
where $L$ is the number of mixture components (clusters), and $p_l$, $\boldsymbol{\mu}_l$, and $\boldsymbol{\Sigma}_l$ are the probability, mean, and covariance matrix, respectively, of mixture component $l$.
The total number of bits available $b_{\text{tot}}$ is $\log_2 M$. 
It is important to note that in both step 1 and step 2, the resulting bit allocations $b_l$ and $b_{l,j}$, respectively, are \emph{not} expected to be integers. The number of levels computed in step 2 corresponding to the bit allocation $b_{l,j}$, that is, $V_{l,j} = 2^{b_{l,j}}$, is ultimately rounded off to the nearest integer, and represents the number of APs. As `bit' generally refers to an integer quantity, to avoid subsequent confusion, we omit the word 'bit' when discussing allocations.  
The codebooks generated in step 3 for each cluster are the required AP locations.

\begin{algorithm}
\caption{PDF Optimized Vector Quantization Algorithm}\label{alg:pdfvq}
\begin{algorithmic}[1]
\State Determine the allocation $b_l$ to cluster $l$ given the total budget $b_{\text{tot}}$ using
\begin{equation}
    2^{b_l} = 2^{b_{\text{tot}}} \frac{ \sqrt{p_l c_l} }{ \sum\limits_{j=1}^L \sqrt{p_l c_l} },
\end{equation}
where $c_l = \sqrt{\lambda_{l,1}\lambda_{l,2}}$, $\boldsymbol{\lambda}_l = \textrm{diag}(\lambda_{l,1},\lambda_{l,2})$, and $\boldsymbol{\Sigma}_l = \mathbf{Q}_l \boldsymbol{\lambda}_l \mathbf{Q}_l^T$ is the eigen value decomposition. 
\State With each cluster $l$, compute the allocation along each dimension $b_{l,j}$, $j = 1, 2$, using
\begin{equation}
    b_{l,j} = \frac{b_l}{2} + \frac{1}{2} \log \left[  \frac{ \lambda_{l,j} }{ c_l }  \right].
\end{equation}
Compute and round off the corresponding level $V_{l,j} = 2^{b_{l,j}}$.
\State Generate the codebook $\mathcal{R}_l$ for each cluster using 
\begin{equation}
    \mathcal{R}_l = \left\{ \mathbf{q} | \mathbf{q} = \mathbf{Q}_l \mathbf{y} + \boldsymbol{\mu}_l, \mathbf{y} \in \mathcal{T}_l \right\},
\end{equation}
where $\mathcal{T}_l$ is the set of vectors given by the Cartesian product $\mathcal{T}_l = \mathcal{S}_{l,1} \times \mathcal{S}_{l,2}$, with $\mathcal{S}_{l,j}$, $j = 1, 2$ being the optimal $V_{l,j}$-level scalar quantizer of a univariate Gaussian with variance $\lambda_{l,j}$.
\end{algorithmic}
\end{algorithm}

\subsection{Gradient Approaches}
\label{ssec:grad_approach}

As alluded to above, the VQ approaches do not explicitly solve the sum rate and minimum rate maximization problems. They, however, provide reasonably good starting points to apply gradient ascent for both problems.
Accordingly, we present both of the gradient calculations below, and are called the \textit{max-sum} and \textit{max-min gradient}, respectively.

\subsubsection{Max-sum gradient}
To maximize sum rate, the gradient update expression with $j$ as the iteration index is
\begin{equation}
    \mathbf{q}_{m}^{(j+1)} = \mathbf{q}_{m}^{(j)} + \delta \frac{ \partial }{ \partial \mathbf{q}_{m}^{(j)} } \left\{ \sum\limits_{k=1}^K \log\left(1 +  \sum\limits_{m=1}^M \beta_{mk} \right) \right\}, \ \forall m,
    \label{eqn:grad_update_eqn_max_sum}
\end{equation}
where in the objective function from \eqref{eqn:sum_rate_max_beta}, we have neglected the shadow fading $z_{mk}$ and replaced the expectation with the sum over the users. The gradient in \eqref{eqn:grad_update_eqn_max_sum} is calculated as
\begin{multline}
   \frac{ \partial }{ \partial \mathbf{q}_{m}^{(j)} } \left\{ \log\left(1 +  \sum\limits_{m=1}^M \beta_{mk} \right) \right\} \\= \frac{\gamma}{2}  \sum\limits_{k=1}^K \frac{1}{1 + \psi_k^{\text{ZF}}} \frac{ (\mathbf{p}_k - \mathbf{q}_{m}^{(j)}) }{ || \mathbf{p}_k - \mathbf{q}_{m}^{(j)} ||^{\gamma+2} }.
\end{multline}

\subsubsection{Max-min gradient}
For minimum rate, taking the gradient of the rate of the worst user is fragile since the absolute value of the minimum rate can vary significantly across the iterations of gradient ascent causing convergence issues. Additionally, we are also interested in evaluating the performance of the cell-free system in terms of the 95\%-likely rate, which quantifies the best rate of the worst 5\% of the users. Accordingly, we consider the sum of rates corresponding to the worst 5\% of the users, represented by $\mathcal{K}_{5\%}$, as
\begin{equation}
    \mathbf{q}_{m}^{(j+1)} = \mathbf{q}_{m}^{(j)} + \delta \frac{ \partial }{ \partial \mathbf{q}_{m}^{(j)} } \left\{ \sum\limits_{k \in \mathcal{K}_{5\%}} \log\left(1 +  \sum\limits_{m=1}^M \beta_{mk} \right) \right\}, \ \forall m,
    \label{eqn:grad_update_eqn_max_min}
\end{equation}
which uses the same simplifications as in \eqref{eqn:grad_update_eqn_max_sum}, and the gradient is
\begin{multline}
   \frac{ \partial }{ \partial \mathbf{q}_{m}^{(j)} } \left\{ \log\left(1 +  \sum\limits_{m=1}^M \beta_{mk} \right) \right\} \\= \frac{\gamma}{2}  \sum\limits_{k \in \mathcal{K}_{5\%}} \frac{1}{1 + \psi_k^{\text{ZF}}} \frac{ (\mathbf{p}_k - \mathbf{q}_{m}^{(j)}) }{ || \mathbf{p}_k - \mathbf{q}_{m}^{(j)} ||^{\gamma+2} }.
\end{multline}

\section{Simulation Methodology and Results}
\label{sec:sim_results}

\subsection{Simulation Parameters}

In a geographical area of dimensions $2 \text{ km} \times 2 \text{ km}$, we consider $M = 32$ APs and $K = 4$ users, since $M \gg K$ for a cell-free system. For the purposes of placement, however, we use $2000$ users that are distributed according to a GMM of the form in \eqref{eqn:gmm}, with covariance $\boldsymbol{\Sigma}_l = \sigma^2_l \mathbf{I}$, where $\sigma_l$ is the standard deviation of mixture component $l$ and $\mathbf{I}$ is the identity matrix.
The GMM used in our simulations has parameters $L=3$, $\boldsymbol{\mu}_1 = [0.5, -0.5]^T$, $\boldsymbol{\mu}_2 = [0, 0.5]^T$, $\boldsymbol{\mu}_3 = [-0.5, 0]^T$, $\sigma_1 = \sigma_2 = \sigma_3 = 100$, $p_1=0.6$, and $p_2=p_3=0.2$.
The pathloss model from \cite[(4.34)]{nay18t} is used with shadow fading ignored for simplicity and the transmit power $\rho_r$ of the users is increased from $5$ to $30$ dB.
%
%
The max-sum and max-min gradient methods utilize step sizes of $\delta = 10^3$ and $\delta = 3 \times 10^4$, respectively.

\subsection{Performance Measures}

The per-user achievable rate is used, and is defined for user $k$ as
\begin{equation}
    R_{k} = \mathbb{E}\left\{ \log_2\left(1 + \psi_{k}^{\text{ZF}}\right) \right\},
\end{equation}
with $\psi_{k}^{\text{ZF}}$ from \eqref{eqn:snr} and the rate values are generated using Monte Carlo iterations. Further, algorithms are run multiple times and the solution that yields the best result is chosen. The maximum number of iterations for the Lloyd algorithm is set to $50$.
For comparison among the algorithms, both the sum rate and the 95\%-likely rate measures are used. The relative performance between algorithms (say, algorithm 2 over algorithm 1) can be calculated by using the following measure expressed as percentage
\begin{equation}
\text{Improvement Ratio} = \frac{ \mathcal{P}^{\text{Algorithm 2}} - \mathcal{P}^{\text{Algorithm 1}} }{ \mathcal{P}^{\text{Algorithm 1}} } \times 100,
\end{equation}
where $\mathcal{P}$ is either the sum rate or 95\%-likely rate.

We note here that while the sum rate  and minimum rate maximization problems are solved by the max-sum and max-min algorithms, respectively, in \cite{nay18a}, we do not include the results here. As described in Section \ref{ssec:sum_snr_maxztn}, the max-sum algorithm does not solve the sum rate problem accurately and generates a colocated-like solution with a remarkably poor 95\%-likely rate performance (reflected in the numerical simulations of \cite{nay18a}). Further, in \cite{gop21}, we compare the grid-based max-min performance with the Lloyd algorithm and show that their rate performances are comparable. Even though a higher 95\%-likely rate is achievable with the max-min algorithm, the simplicity of the Lloyd algorithm outweighs the complexity associated with the increased grid resolution needed.

\subsection{Numerical Results}

\begin{figure}[t!]
	\centering
	\includegraphics [scale=0.555] {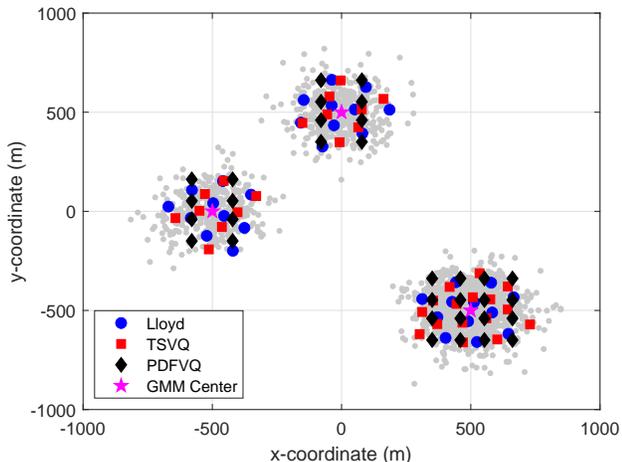}
	\caption{Final AP locations of the Lloyd algorithm, TSVQ, and PDFVQ at $\rho_r = 30$ dB.}
	\label{fig:tsvq_pdfvq_ap_locs}
\end{figure}
\begin{figure}[t!]
	\centering
	\includegraphics [scale=0.555] {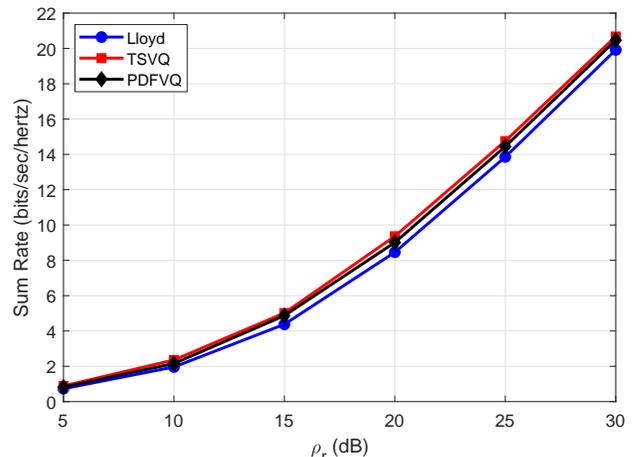}
	\caption{Sum rate as a function of $\rho_r$ for the Lloyd algorithm, TSVQ, and PDFVQ.}
	\label{fig:tsvq_pdfvq_sum_rate}
\end{figure}
\begin{figure}[t!]
	\centering
	\includegraphics [scale=0.555] {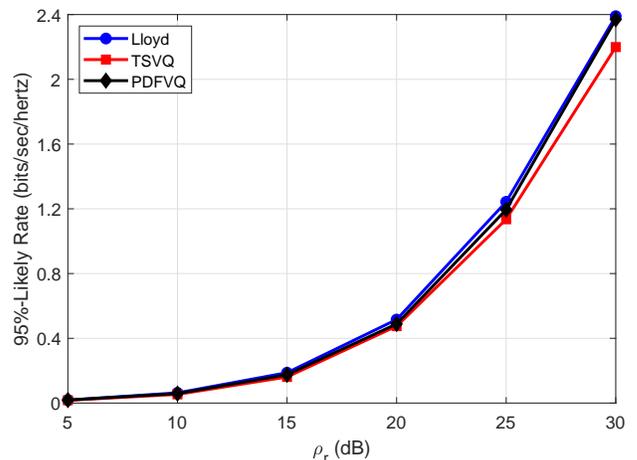}
	\caption{95\%-likely rate as a function of $\rho_r$ for the Lloyd algorithm, TSVQ, and PDFVQ.}
	\label{fig:tsvq_pdfvq_95_rate}
\end{figure}
\begin{table}[t!]
\renewcommand{\arraystretch}{1.3}
\caption{Rate Improvement of TSVQ and PDFVQ Relative to the Lloyd Algorithm at $\rho_r = 30$ dB}
\label{tab:improv_tsvq_pdfvq}
\centering
\begin{tabular}{|c|c|c|}
\hline
Algorithm & Sum Rate & 95\%-Likely Rate\\
\hline
TSVQ & $3.82$\% & $-7.98$\% \\
PDFVQ & $2.78$\% & $-0.82$\% \\
\hline
\end{tabular}
\end{table}

\textit{Experiment 1.}
We compare the performances of TSVQ and PDFVQ to the standard Lloyd algorithm. 
Their final AP locations and the distribution of the 2000 users (as gray circles) are shown in Fig. \ref{fig:tsvq_pdfvq_ap_locs}, where the Lloyd algorithm results in APs that are more distributed than TSVQ. This is expected since the TSVQ algorithm successively splits Voronoi regions and after completion does not necessarily associate each user to its closest AP. For PDFVQ, the allocation procedure within each cluster results in 3.85, 2.93, and 2.93 APs along each dimension, respectively, since the GMM clusters are symmetric along both x- and y-coordinates. Although rounding off these values would result in 4, 3, and 3 APs along each coordinate, the total number of APs would then be 34. Thus, to limit to $M = 32$ APs, we select 4, 2, and 2 APs along the x-coordinate and 4, 4, and 4 APs along the y-coordinate. This combination is selected as it provides the best result through repeated trials. The AP locations are observed to be more regular and grid-like due to the scalar quantizers used in the transform coding design, when compared to the standard Lloyd algorithm or TSVQ.
The sum and 95\%-likely rates of the three VQ-based algorithms are plotted in Fig. \ref{fig:tsvq_pdfvq_sum_rate} and Fig. \ref{fig:tsvq_pdfvq_95_rate}, respectively. The improvements in sum and 95\%-likely rates, expressed as percentages, of TSVQ and PDFVQ over the Lloyd algorithm at transmit power $\rho_r = 30$ dB are noted in Table \ref{tab:improv_tsvq_pdfvq}. It is seen that while the sum rate of TSVQ is greater than that of the Lloyd algorithm (by nearly $4$\%), the 95\%-likely rate is worse (by nearly $8$\%). For PDFVQ, the sum rate offered is close to that of TSVQ while the 95\%-likely rate approaches but does not equal that of the Lloyd algorithm (with the difference being less than $1$\%), due to the fact that the AP positions are far away from a small group of users in two mixture components. While for the user distribution considered PDFVQ offers a solution providing higher sum rate and similar 95\%-likely rate to the Lloyd algorithm, it can be expected that this performance will reduce as the variance of the mixture components is increased.

\begin{figure}[t!]
	\centering
	\includegraphics [scale=0.555] {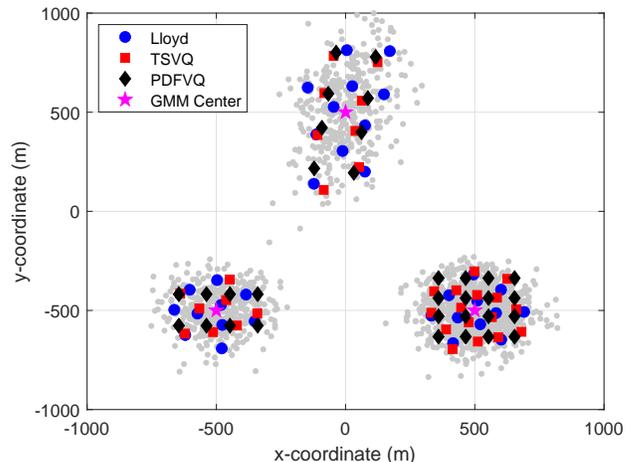}
	\caption{Final AP locations of the Lloyd algorithm, TSVQ, and PDFVQ with a full covariance matrix at $\rho_r = 30$ dB.}
	\label{fig:tsvq_pdfvq_nondiag_ap_locs}
\end{figure}

\begin{figure}[t!]
	\centering
	\includegraphics [scale=0.555] {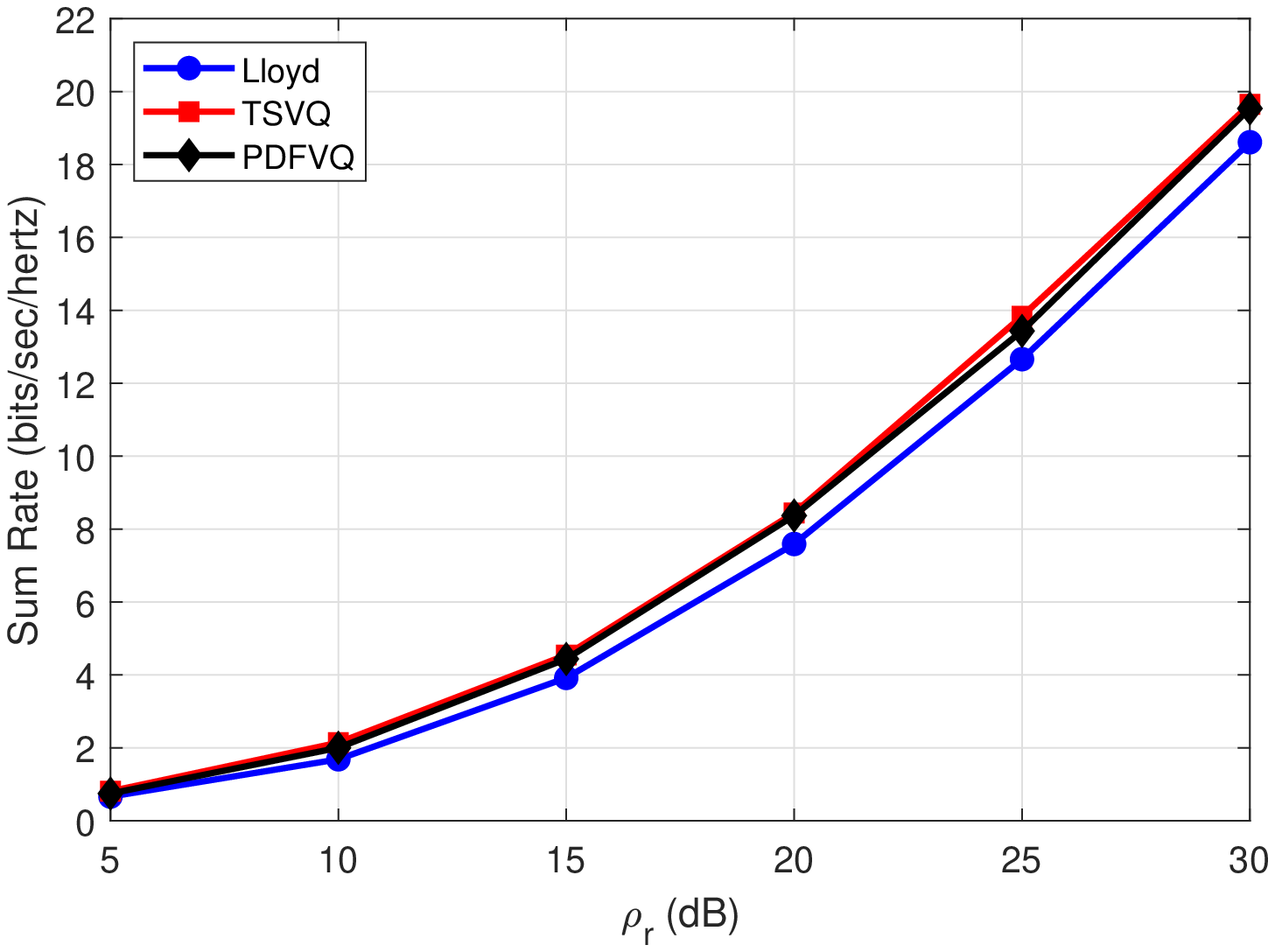}
	\caption{Sum rate as a function of $\rho_r$ for the Lloyd algorithm, TSVQ, and PDFVQ with a full covariance matrix.}
	\label{fig:tsvq_pdfvq_nondiag_sum_rate}
\end{figure}

\begin{figure}[t!]
	\centering
	\includegraphics [scale=0.555] {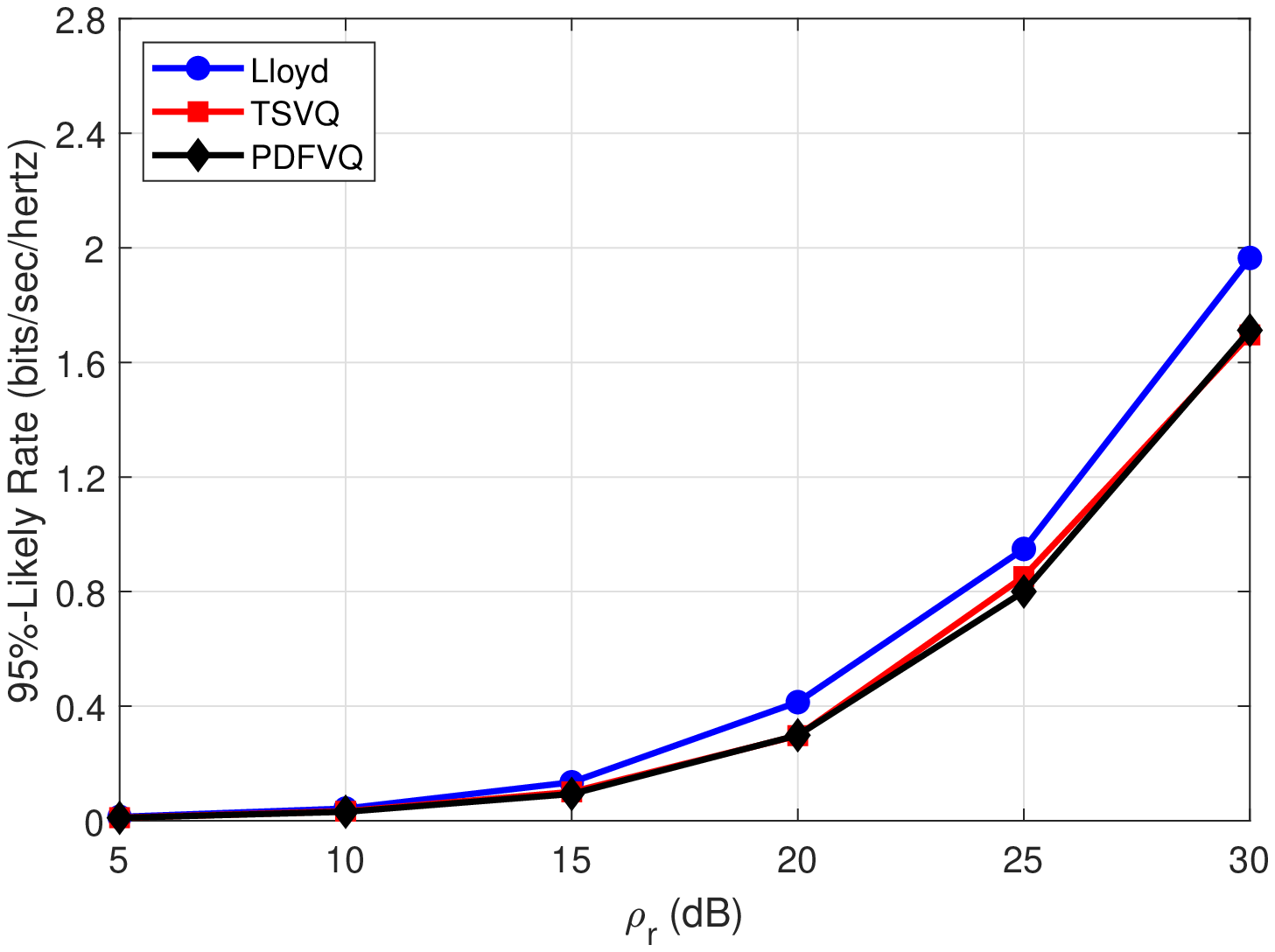}
	\caption{95\%-likely rate as a function of $\rho_r$ for the Lloyd algorithm, TSVQ, and PDFVQ with a full covariance matrix.}
	\label{fig:tsvq_pdfvq_nondiag_95_rate}
\end{figure}

\textit{Experiment 2.}
In Experiment 1, we compared the VQ approaches for a GMM consisting of Gaussian mixture components with spherical (proportionate to the identity matrix) covariance matrices. In order to show an example of PDFVQ applied to a full (non-diagonal) covariance matrix, we now consider a user distribution where the second component of the GMM considered is modified to have a covariance matrix of
$\boldsymbol{\Sigma}_2 = \begin{bsmallmatrix}
    \sigma^2 & 2\sigma^2/3\\
    2\sigma^2/3 & 2\sigma^2
    \end{bsmallmatrix}$
with $\sigma = 100$. When PDFVQ is applied to this distribution, $4$, $2$, $3$ APs along the x-coordinate and $4$, $5$, and $3$ APs along the y-coordinate are allocated, totaling $35$ APs. The best allocation for the desired $32$ APs is found to be $4$, $2$, and $4$, and $4$, $4$, and $2$ APs along the x- and y-coordinates, respectively. The positions of such APs along with those of the Lloyd algorithm and TSVQ are shown in Fig. \ref{fig:tsvq_pdfvq_nondiag_ap_locs}. The sum rate and 95\%-likely rates corresponding to these locations are provided in Fig. \ref{fig:tsvq_pdfvq_nondiag_sum_rate} and Fig. \ref{fig:tsvq_pdfvq_nondiag_95_rate}, respectively, and the rate improvements are tabulated in Table \ref{tab:improv_tsvq_pdfvq_nondiag}. Similar to the GMM with spherical Gaussians in Experiment 1, we observe that PDFVQ is able to match the sum rate performance of TSVQ. However, unlike above, the 95\%-likely rate, like TSVQ, is lower than the Lloyd algorithm by nearly $13$\% since the space-filling advantage of the Lloyd algorithm is lost. 
Based on the above two experiments, it could be concluded that PDFVQ is a reasonable alternative to TSVQ that provides a similar or superior performance.

Although it is not the case here, we also note that the clusters of the GMM may be placed quite close to one another or may even overlap marginally. When PDFVQ is applied, the APs associated with the overlapping clusters may be closely placed and is inefficient. In such a scenario, starting from the cluster pair with most overlap (the degree of which can be computed by a measure such as overlap rate \cite{sun11}), the AP allocation of the cluster with the higher 95\%-likely rate could be decreased and re-allocated to another cluster with a poor performance. This process can be repeated for all overlapping clusters to improve the 95\%-likely rate of the network.

\begin{table}[t!]
\renewcommand{\arraystretch}{1.3}
\caption{Rate Improvement of TSVQ and PDFVQ Relative to the Lloyd Algorithm With a Full Covariance Matrix at $\rho_r = 30$ dB}
\label{tab:improv_tsvq_pdfvq_nondiag}
\centering
\begin{tabular}{|c|c|c|}
\hline
Algorithm & Sum Rate & 95\%-Likely Rate\\
\hline
TSVQ & $5.59$\% & $-13.67$\% \\
PDFVQ & $4.98$\% & $-12.85$\% \\
\hline
\end{tabular}
\end{table}

\begin{figure}[t!]
	\centering
	\includegraphics [scale=0.555] {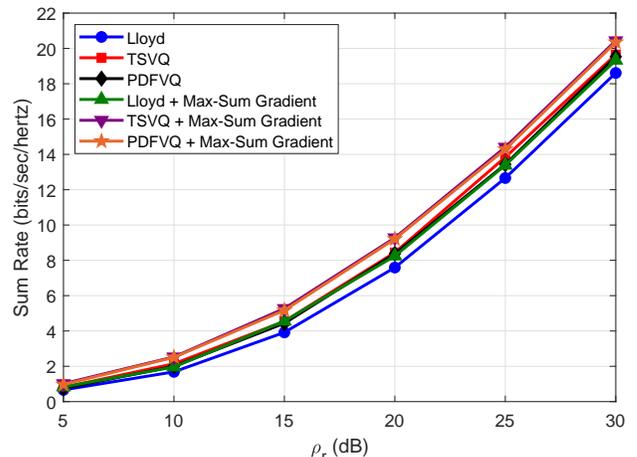}
	\caption{Sum rate as a function of $\rho_r$ for the Lloyd algorithm, TSVQ, and PDFVQ with a full covariance matrix along with max-sum gradient.}
	\label{fig:max_sum_grad_sum_rate}
\end{figure}

\begin{figure}[t!]
	\centering
	\includegraphics [scale=0.555] {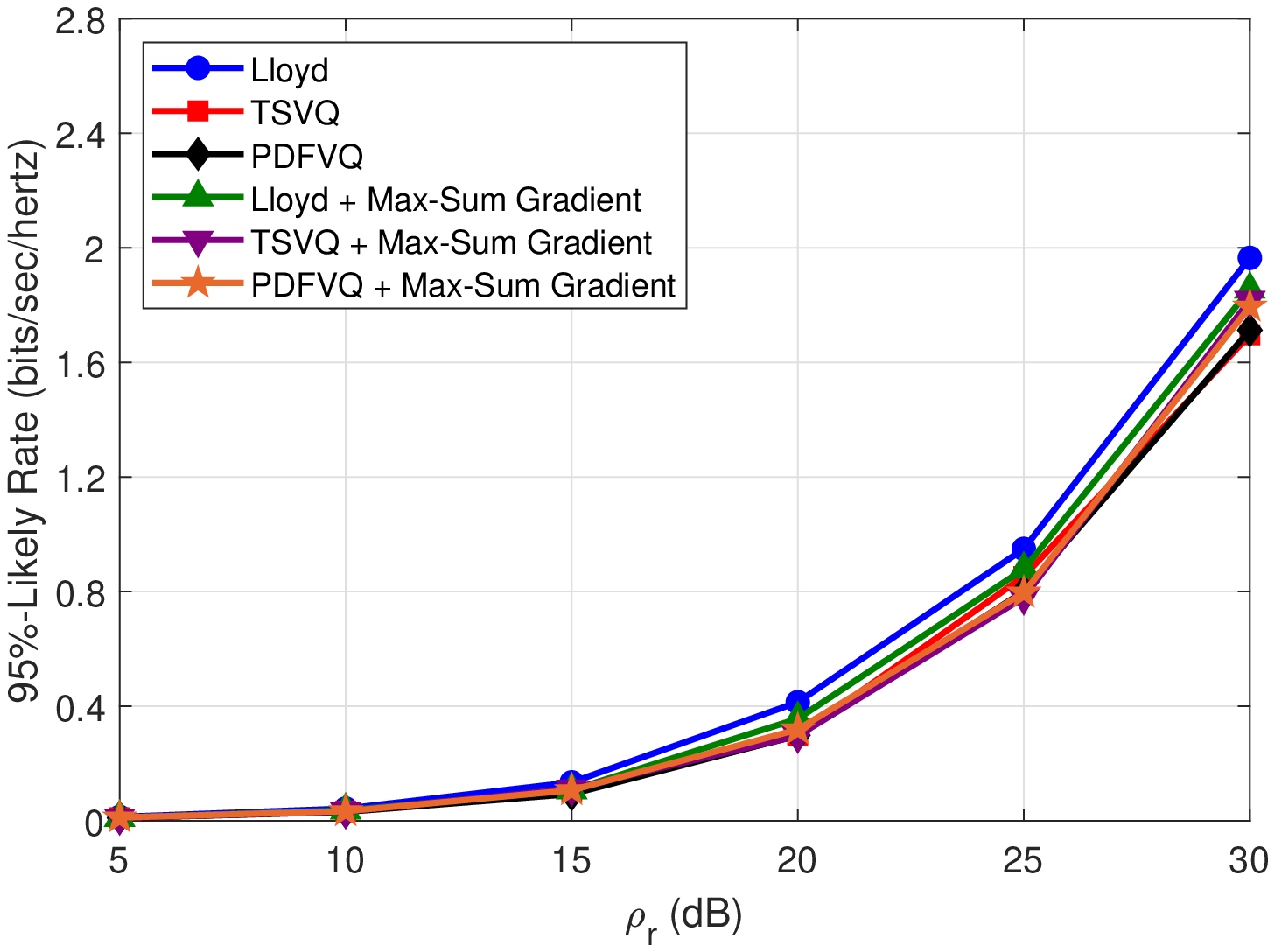}
	\caption{95\%-likely rate as a function of $\rho_r$ for the Lloyd algorithm, TSVQ, and PDFVQ with a full covariance matrix along with max-sum gradient.}
	\label{fig:max_sum_grad_95_rate}
\end{figure}

\begin{figure}[t!]
	\centering
	\includegraphics [scale=0.555] {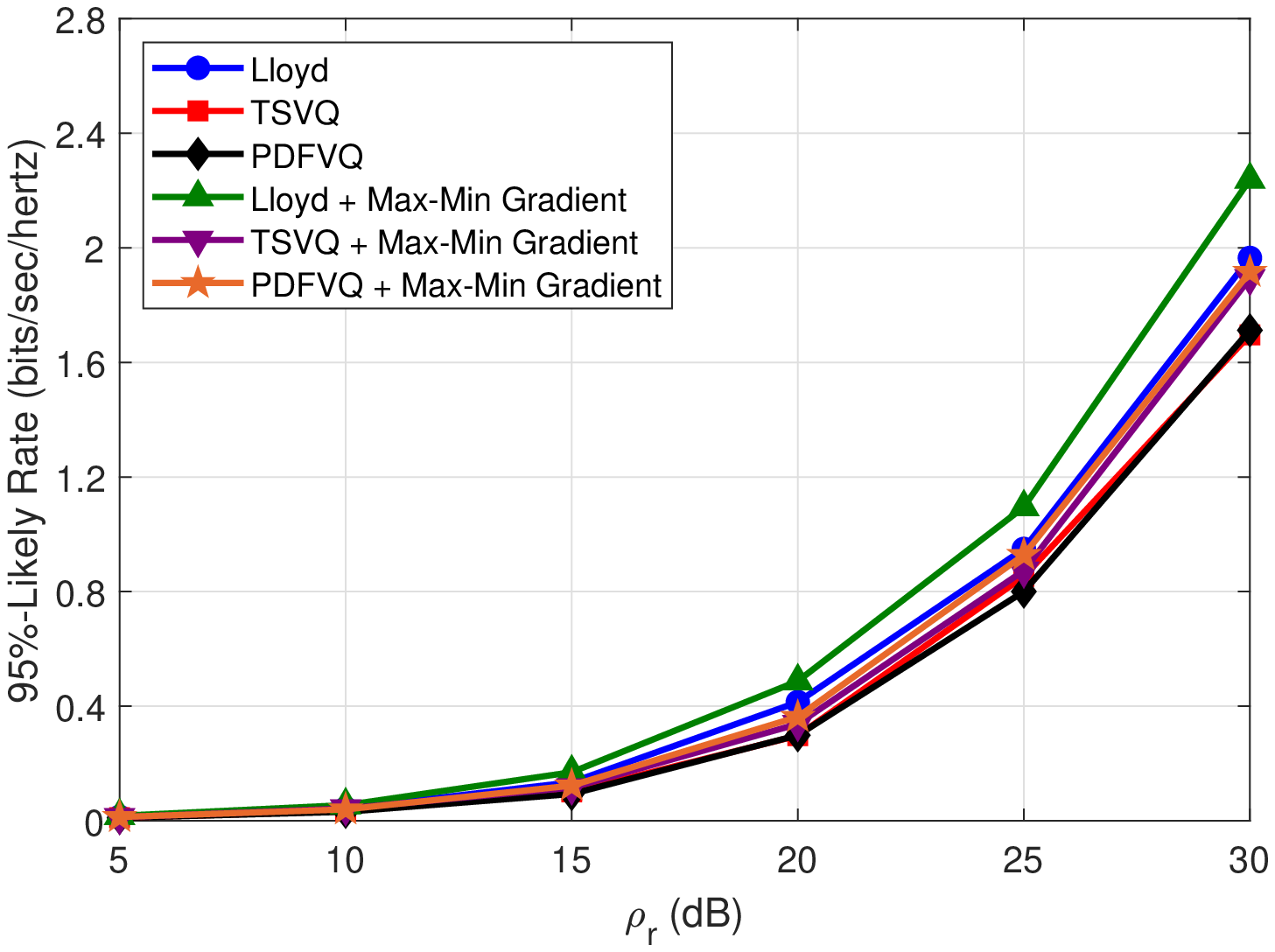}
	\caption{95\%-likely rate as a function of $\rho_r$ for the Lloyd algorithm, TSVQ, and PDFVQ with a full covariance matrix along with max-min gradient.}
	\label{fig:max_min_grad_95_rate}
\end{figure}

\begin{figure}[t!]
	\centering
	\includegraphics [scale=0.555] {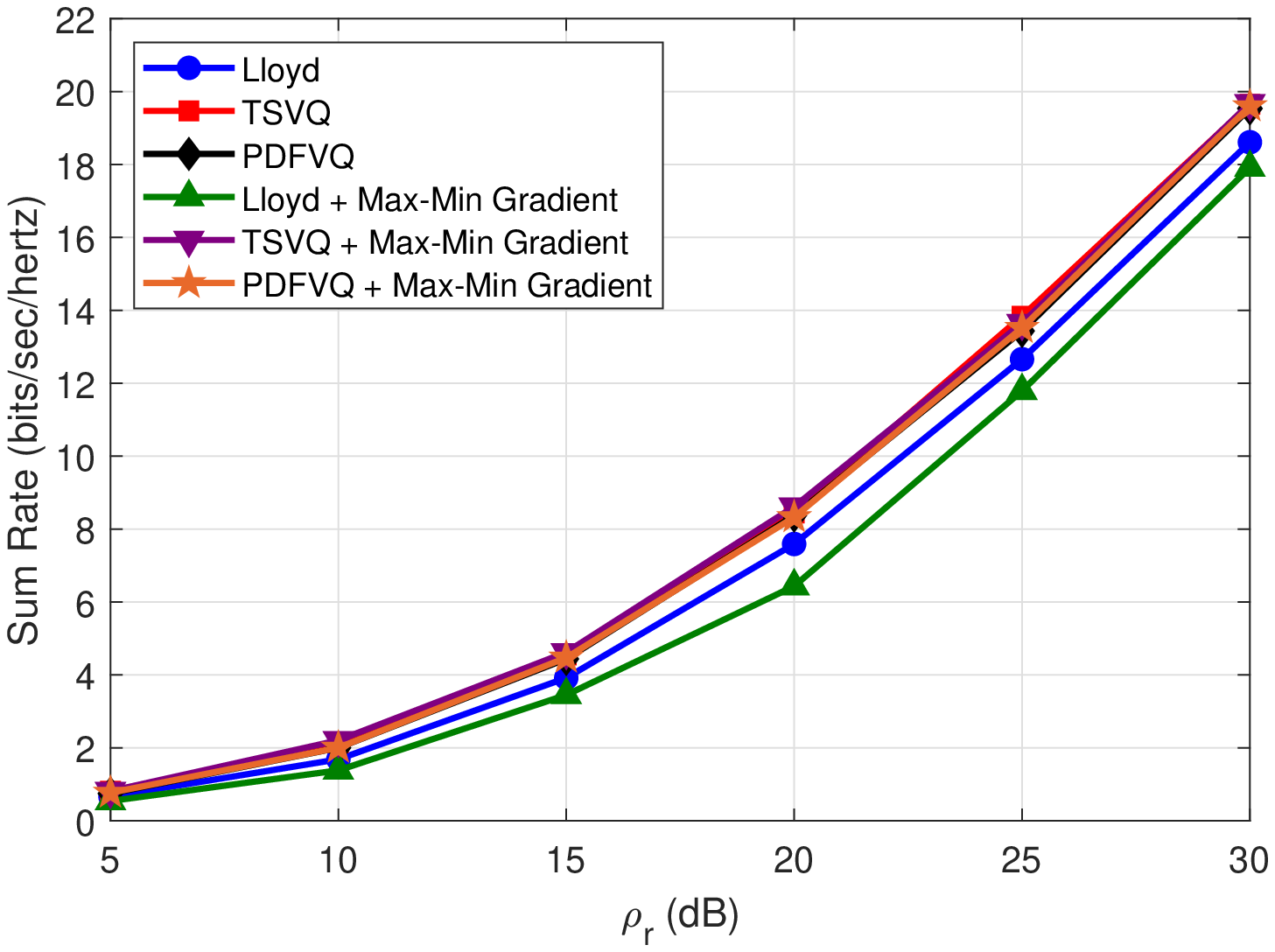}
	\caption{Sum rate as a function of $\rho_r$ for the Lloyd algorithm, TSVQ, and PDFVQ with a full covariance matrix along with max-min gradient.}
	\label{fig:max_min_grad_sum_rate}
\end{figure}

\textit{Experiment 3.}
In this experiment, we use the gradient approaches outlined in Section \ref{ssec:grad_approach} to improve the sum rate and 95\%-likely rate performances shown in Experiment 2 for the GMM with a full covariance matrix. 
First, we show that using the max-sum gradient ascent, we can increase the sum rates of each of the Lloyd, TSVQ, and PDFVQ performances, shown in Fig. \ref{fig:max_sum_grad_sum_rate}. Table \ref{tab:improv_maxsum_grad} lists the percentage rate improvements of the max-sum gradient applied to each VQ approach, over the Lloyd algorithm alone. The gradient ascent applied to both the TSVQ and PDFVQ AP locations yields the highest sum rate (nearly $10$\% over the Lloyd algorithm) while the rate obtained when the ascent operation is applied with the Lloyd algorithm as the starting point is nearly the same as the sum rates of PDFVQ and TSVQ. This occurrence is due to the fact that the gradient ascent iterations with the Lloyd AP solutions as the initial points converge to a local optimum which is different from that obtained when the ascent is applied to PDFVQ or TSVQ.
The 95\%-likely rate performances shown in Fig. \ref{fig:max_sum_grad_95_rate}, when the max-sum gradient is applied to PDFVQ and TSVQ, do not change significantly (an increase is observed at $\rho_r = 30$ dB) and decrease when the gradient is applied to the Lloyd AP positions. Thus, in terms of sum rate, the PDFVQ provides the best solution out of all VQ approaches and a further increase in sum rate (about $4$\% over PDFVQ) without negatively affecting the minimum rate performance (about $5$\% increase) is achieved by using the max-sum gradient ascent.
Next, the max-min gradient ascent is applied and the 95\%-likely rate is observed to increase as shown in Fig. \ref{fig:max_min_grad_95_rate}. The best rate is obtained when the ascent algorithm is applied to the Lloyd solution as opposed to when it is applied to either TSVQ or PDFVQ where it is able to match the performance of the Lloyd algorithm. Table \ref{tab:improv_maxmin_grad} informs us that while a $14$\% improvement in the 95\%-likely rate is achieved by applying the gradient to the Lloyd solution, the difference of the gradient applied to TSVQ or PDFVQ from the Lloyd algorithm is only up to $4$\%. The sum rates corresponding to the 95\%-likely rates are plotted in Fig. \ref{fig:max_min_grad_sum_rate}, where as a result of the 95\%-likely rate improvement, the sum rate when the ascent is applied to the Lloyd algorithm is the least value. In contrast, despite the increase in 95\%-likely rate, the sum rate performances when ascent is applied to TSVQ and PDFVQ are nearly the same as TSVQ and PDFVQ itself. Hence, with its simpler design, PDFVQ along with max-min gradient offers a good tradeoff between a 95\%-likely rate similar to and a sum rate higher (by over $5$\%) than the Lloyd algorithm.
It should, however, be mentioned that if 95\%-likely rate (or minimum rate) is the sole performance measure of interest, the Lloyd algorithm alone is a straightforward choice. The gradient approach requires the choice of an appropriate step size for convergence while the Lloyd algorithm is known to converge \cite{sab86}.

\begin{table}[t!]
\renewcommand{\arraystretch}{1.3}
\caption{Rate Improvement of the VQ Approaches with the Max-Sum Gradient Relative to the Lloyd Algorithm at $\rho_r = 30$ dB}
\label{tab:improv_maxsum_grad}
\centering
\begin{tabular}{|c|c|c|}
\hline
Algorithm & Sum Rate & 95\%-Likely Rate\\
\hline
Lloyd + Max-Sum Gradient & $3.88$\% & $-5.60$\% \\
TSVQ + Max-Sum Gradient & $9.80$\% & $-7.52$\% \\
PDFVQ + Max-Sum Gradient & $9.34$\% & $-8.66$\% \\
\hline
\end{tabular}
\end{table}

\begin{table}[t!]
\renewcommand{\arraystretch}{1.3}
\caption{Rate Improvement of the VQ Approaches with the Max-Min Gradient Relative to the Lloyd Algorithm at $\rho_r = 30$ dB}
\label{tab:improv_maxmin_grad}
\centering
\begin{tabular}{|c|c|c|}
\hline
Algorithm & 95\%-Likely Rate & Sum Rate\\
\hline
Lloyd + Max-Min Gradient & $13.93$\% & $-3.73$\% \\
TSVQ + Max-Min Gradient & $-3.60$\% & $5.73$\% \\
PDFVQ + Max-Min Gradient & $-2.48$\% & $5.30$\% \\
\hline
\end{tabular}
\end{table}

\begin{figure}[t!]
     \centering
     \begin{subfigure}[t]{0.49\columnwidth}
         \centering
         \includegraphics[scale=0.28]{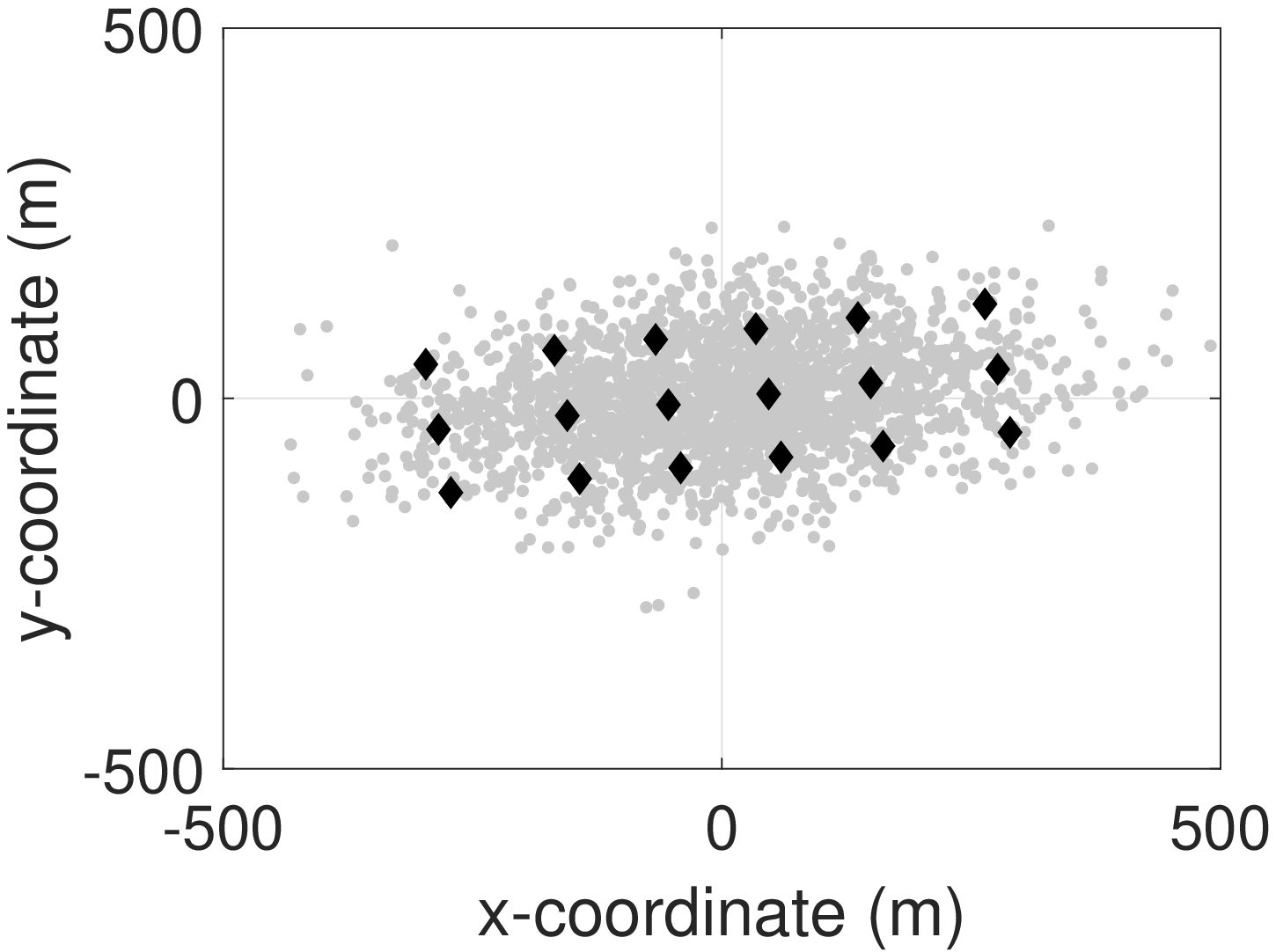}
         \caption{Density A.}
         \label{fig:cf_paper_adap_conf_A}
     \end{subfigure}%
     \begin{subfigure}[t]{0.49\columnwidth}
         \centering
         \includegraphics[scale=0.28]{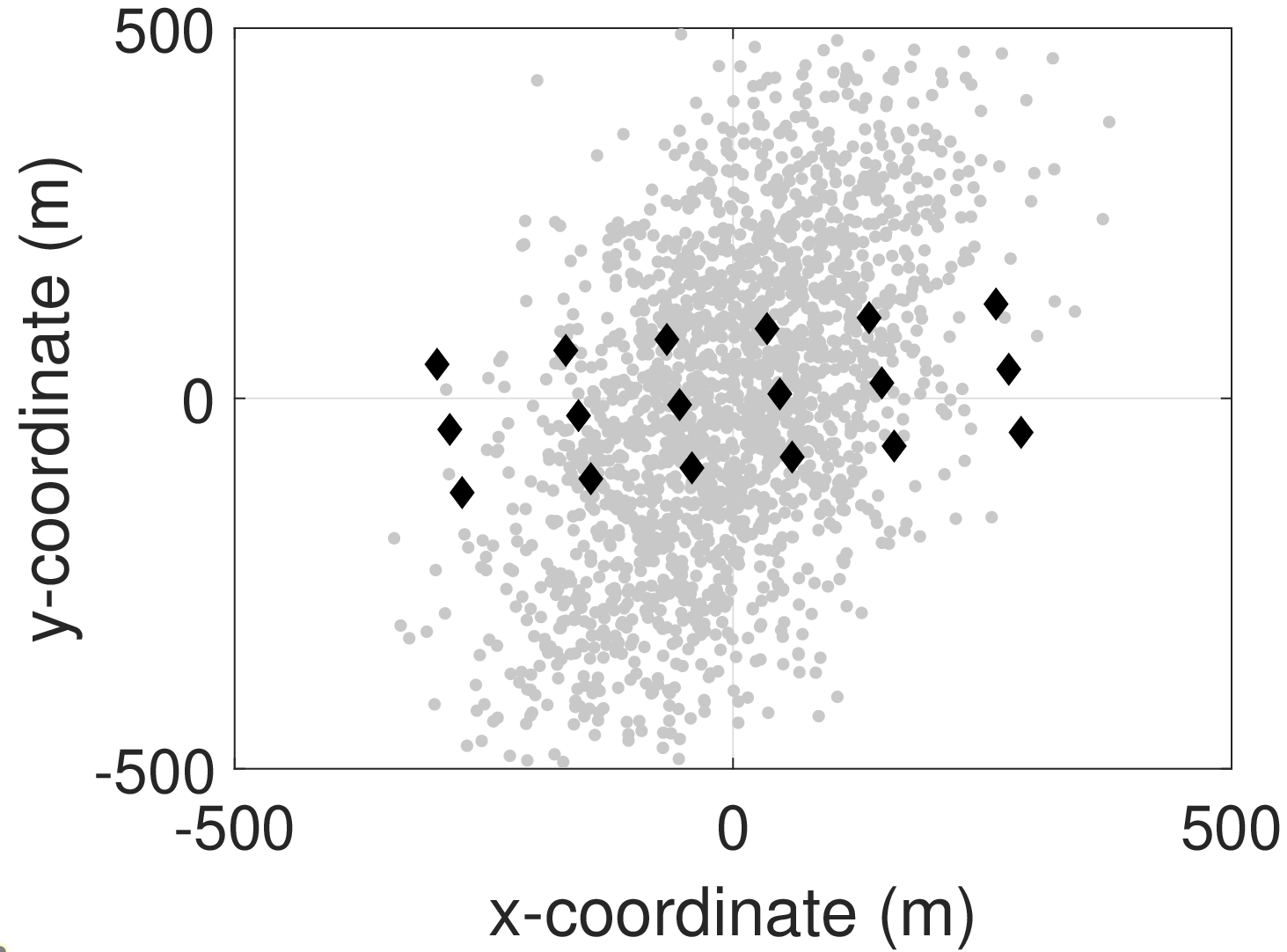}
         \caption{Density B.}
         \label{fig:cf_paper_adap_conf_B}
     \end{subfigure}
        \caption{The two user densities A and B considered for Experiment 4 along the PDFVQ AP locations matched to density A.}
        \label{fig:cf_paper_adap_confs}
\end{figure}

\begin{figure}[t!]
	\centering
	\includegraphics [scale=0.555] {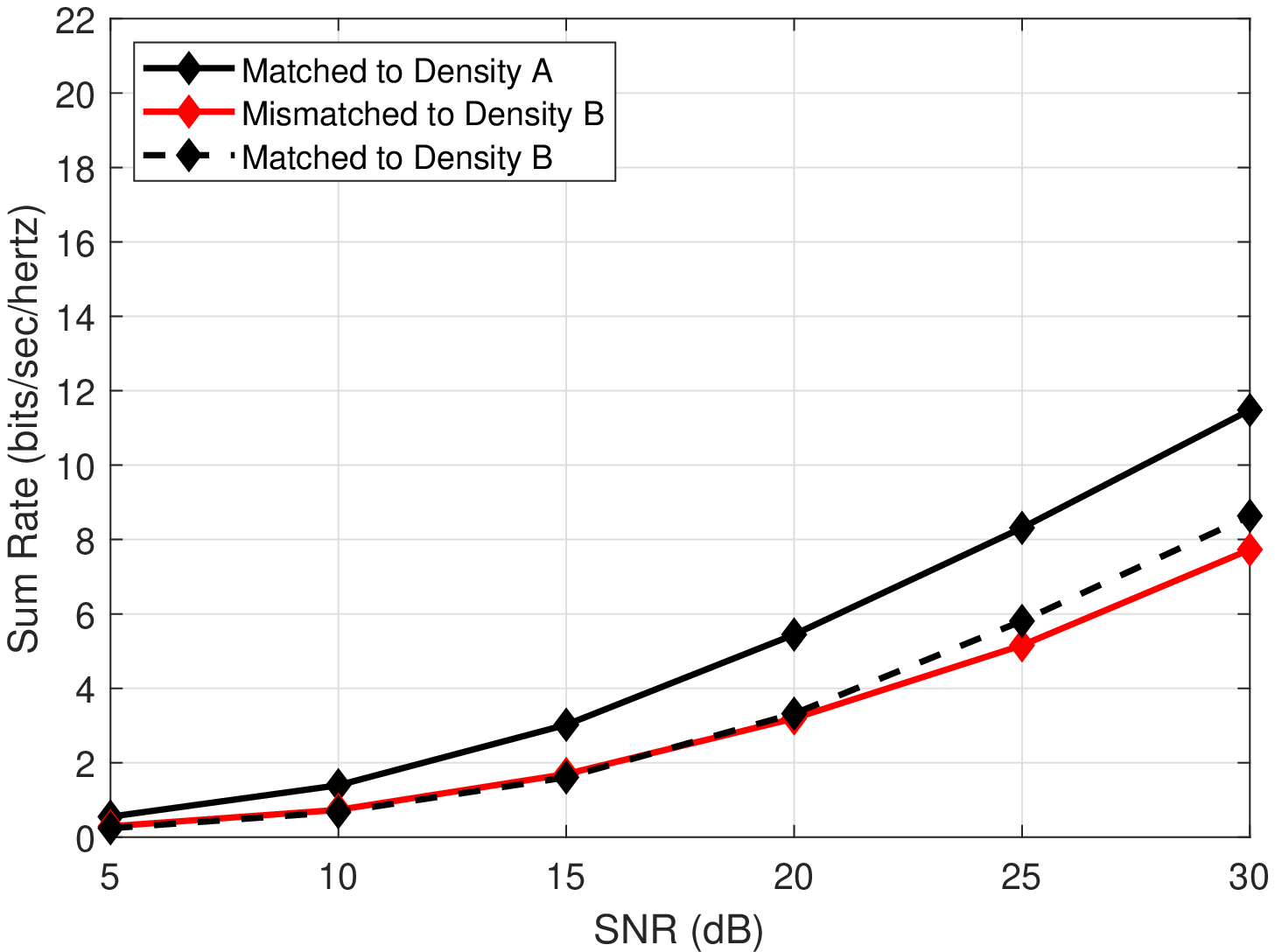}
	\caption{Sum rate as a function of $\rho_r$ for Experiment 4.}
	\label{fig:cf_paper_adap_sum_rate}
\end{figure}

\begin{figure}[t!]
	\centering
	\includegraphics [scale=0.555] {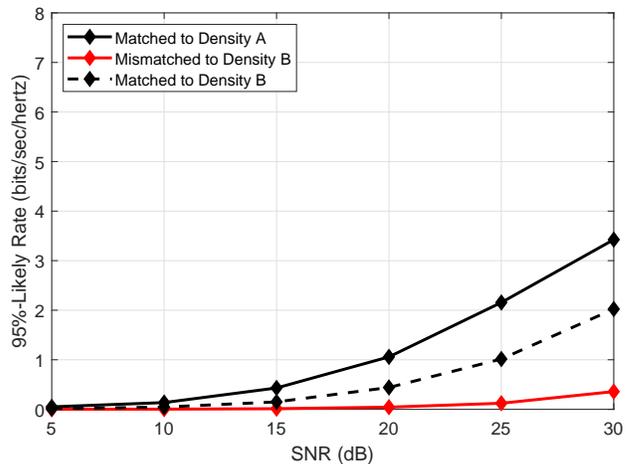}
	\caption{95\%-likely rate as a function of $\rho_r$ for Experiment 4.}
	\label{fig:cf_paper_adap_95_rate}
\end{figure}

\textit{Experiment 4.}
In this experiment, we quantify the effect of a time-varying user density on the network performance and show the need for an easily adaptable AP placement algorithm. We consider a simple situation where the users are \emph{initially} distributed as a single-cluster GMM ($L = 1$ in \eqref{eqn:gmm}) with mean $\boldsymbol{\mu} = [0,0]^T$ and covariance $\boldsymbol{\Sigma} = \begin{bsmallmatrix}
\sigma^2 & \sigma^2/3\\
\sigma^2/3 & \sigma^2/2
\end{bsmallmatrix}$
where $\sigma = 200$, and call it density A. Assuming a total of $M = 18$ APs, we calculate the PDFVQ AP locations for this user density. Over time, we consider that the user distribution changes to density B, where the users are more spread out along another direction to that of density A, with covariance matrix $\boldsymbol{\Sigma} = \begin{bsmallmatrix}
\sigma^2/2 & \sigma^2/2\\
\sigma^2/2 & \sigma^2
\end{bsmallmatrix}$
and $\sigma = 300$. These two user densities and the PDFVQ AP locations determined for density A are shown in Fig. \ref{fig:cf_paper_adap_confs}.
We evaluate the sum rate and 95\%-likely rate performances when the AP locations are matched to density A and consider both user densities A and B, so that the loss due to mismatch of the AP locations to the user density B is also shown. For completeness, we also compute the rate values when PDFVQ AP locations are determined and matched to density B. As expected, in both the sum rate plotted in Fig. \ref{fig:cf_paper_adap_sum_rate} and the 95\%-likely rate plotted in Fig. \ref{fig:cf_paper_adap_95_rate}, there is a significant loss in performance for the PDFVQ AP locations which are matched to density A, when the users re-position to density B. The relative losses in the sum rate and 95\%-likely rate at $\rho_r = 30$ dB are $32.67$\% and $89.54$\%, respectively. To prevent this performance decrease, there is a need to re-calculate the AP locations for the new user density. Clearly, when the AP locations are matched via PDFVQ to the user density B, the rate performances in Fig. \ref{fig:cf_paper_adap_sum_rate} and Fig. \ref{fig:cf_paper_adap_95_rate} are improved over the diminished performance of the mismatched case. With easy adaptability and the other aforementioned advantages, PDFVQ offers the best method for cell-free AP placement among the VQ techniques discussed in this paper. In future work, we will address how the changing user densities can be learned for use in PDFVQ.


\section{Conclusion}
\label{sec:concl}

In this paper, we have addressed access point (AP) placement in cell-free massive multiple-input-multiple-output (MIMO) systems under a throughput criteria. We investigated the two main optimization problems in this regard, namely the sum rate and minimum rate maximization problems. To understand their solution frameworks, simple examples were constructed and analyzed exposing the difficulty in solving the problems.
Therefore, as a practical approach, the use of vector quantization (VQ)-based methods, namely the popular Lloyd algorithm, tree-structured VQ (TSVQ), and probability density function optimized VQ (PDFVQ), to cell-free AP placement, was investigated. Among the three algorithms presented, although the tree-structured VQ (TSVQ) provides better sum rate (as it fosters cooperation among APs by placing them closer) compared to the Lloyd algorithm, it suffers from high complexity, poor scalability, and the inability to easily adapt to new environments. PDFVQ, which overcomes the aforementioned shortcomings, allowed a more efficient generation of the codebook and generated a sum rate similar to and 95\%-likely rate higher than TSVQ and close to the Lloyd algorithm. Additionally, for gradient-based maximization methods, PDFVQ is found to provide good initial points. It was observed numerically that, over the Lloyd algorithm, an increase of $9$\% in sum rate and a difference of just $2.5$\% in the 95\%-likely rate was achieved by applying max-sum and max-min gradient ascent algorithms, respectively, with the PDFVQ AP locations as starting points. Thus, PDFVQ offers a convenient, less computationally intensive, and easily scalable AP placement technique for cell-free networks.

\appendices




%
%
\bibliographystyle{IEEEtran}
\bibliography{references}

%




\end{document}